# Virtual Scanner Games: expanding access to Magnetic Resonance (MR) education through interactive web tutorials

Gehua Tong[1], Rishi Ananth[2], John Thomas Vaughan, Jr.[1,3], Sairam Geethanath[3,4,*]

[1]Biomedical Engineering, Columbia University, New York, NY, United States
[2]College of Arts and Sciences, University of Washington, Seattle, WA, United States
[3]Columbia Magnetic Resonance Research Center, Columbia University, New York, NY, United States
[4]Accessible MR Laboratory, BioMedical Engineering and Imaging Institute, Dept. of Diagnostic, Molecular and Interventional Radiology, Icahn School of Medicine at Mt. Sinai, New York, NY, United States
[*]Corresponding author

**Abstract**

*Background:* Magnetic Resonance Imaging (MRI) is a medical imaging technique that combines principles of physics, math, and engineering to look inside humans and animals, hence delivering a lifesaving technology. However, resource-poor regions aiming to expand access to MRI suffer from a lack of imaging expertise. Education of a new generation of MRI technicians and researchers in these regions is needed to make this technology more equitable and sustainable around the world.

*Results:* We developed Virtual Scanner Games, an open-source web application that allows students to explore fundamental concepts in MRI. Four modules illustrated imaging basics, imaging biology, imaging physics, and imaging design. Feedback was collected from high school and early college students along with faculty, postdoctoral fellows, and doctoral students working in the field of MRI.

*Conclusions:* We have shown a collection of games that can be used independently by high school students and people with no imaging background; they could also be used as demonstration tools in formal courses. Further assessment of their educational value through longer term usage is needed.

**Introduction**

Magnetic Resonance Imaging (MRI) is a medical imaging technique that provides high spatial resolution and multiple contrasts. It is used extensively for advancing science and clinical practice with applications ranging from looking at brain activity and white matter structure to imaging porous rocks and mummies (Bakhmutov, 2011; Le Bihan et al., 2001; Ogawa et al., 1990; Shin et al., 2010). Clinically, it serves key purposes including diagnosis, monitoring, and surgery assistance in an non-invasive manner. However, access to MRI in the world is highly unequal (Geethanath & Vaughan Jr., 2019). In 2016, Nigeria had the highest number of MRI machines per million population at 0.30 among 16 surveyed West African countries; in comparison, the median number for twelve countries with gross domestic product > 10000 was 9.76 (Ogbole et al., 2018). The global average was about 7 per million population in 2020 (Y. Liu et al., 2021). While having more scanners is necessary for expanding access, scanner numbers alone do not help sustain operations: local expertise is required to maintain the scanner's productivity in the long run (Anazodo et al., 2023). As one solution, open-source and free educational software may help teach future generations of MRI technicians and scientists and is recognized as a key need, as reflected in the 2019 International Society for Magnetic Resonance Imaging in Medicine Junior Fellow African Challenge topic (*2019 Junior Fellow Challenge: Africa – ISMRM*, n.d.).

Non-profit organizations such as RAD-AID support the expansion of access to medical imaging through local assessment of readiness, installation of imaging equipment, and education of personnel (Mollura et al., 2017). The educational aspect is based on hands-on training at the targeted location followed by an online learning center. To complement this, there is a need for simulation-based software that allow new practitioners and students to practice running scans and learn imaging physics interactively. Many existing open-source Magnetic Resonance (MR) simulation programs provide functionality with educational goals or simulation capabilities helpful for teaching MRI concepts (Benoit-Cattin et al., 2005; Hackländer & Mertens, 2005; F. Liu et al., 2017; Pizetta, 2018; Stöcker et al., 2010; Tong et al., 2019; Treceño-Fernández et al., 2019; Wilhjelm et al., 2018). Some focus on emulating the scanning procedure; others, numerical simulation based on the Bloch equations which model physical processes underlying MR signals.

One-dimensional to three-dimensional Images are generated either with signal models or simulated data. The user may set scanning parameters, look at output images and diagrams, or perform other manipulations, such as those on k-space, to gain understanding of the imaging process. These tools are summarized in Table 1.

On the other hand, custom MR hardware systems for education have also been developed to help students gain real scanner operating experience and an understanding of the MR hardware components that work together (Cooley et al., 2020; Niumag, n.d.; Pure Devices, n.d.; SelectScience, n.d.; Twieg et al., 2013; Wright et al., 2001). These scanners have magnetic field strengths from 0.2 to 0.5T and are available either as commercial products or open hardware descriptions. They often have associated educational materials and console software. A selection of these systems are summarized in Table 2.

Each of these platforms holds the assumption that MR education starts at the college or graduate level. To reach those levels or be enrolled in those courses, students would need to be exposed to the technology or related fields beforehand. Therefore, a gap exists in populations that are younger or lack access to formal courses provided by an advanced educational infrastructure. MR is a multifaceted field and includes many aspects of mathematical, biological, physical, and spatial thinking. Many of its concepts and processes need visualization. The flourishing of open-source software and open hardware in MR enables the development of new educational possibilities (Open Source Imaging Initiative, n.d.). Furthermore, combining open-source software and hardware may help expand access and lower costs of hands-on education.

Therefore, a platform aimed at students at or above the educational background of a US high-schooler should exist to equip future engineers and radiologists in underserved geographies with tools to initiate their MR education and to supplement existing high school Science, Technology,

Engineering, Art, and Math (STEAM) curricula in the US with a specific real-world application. To this end, we developed the Virtual Scanner Games, which consists of eight interactive web game pages. The setup includes an application that can be used as a stand-alone tutorial or with accompanying lab manuals in a traditional classroom setting. It is also possible to extend the application so it controls hardware for the hands-on portion of the education. In the following sections, we detail the design and open-source implementation of the games and summarize feedback from testing and web deployment.

**Methods**

*Software components*

The games were implemented as a zero-footprint web browser application using Python 3.9 and packages including Flask 2.1 for web framework, Bootstrap 5 for styling, JQuery 3.3 for frontend interactivity, and Plotly 5.9 for image and data displays. The versions of these softwares and libraries are specified in the requirements file as part of the GitHub repository (imr-framework, 2022/2022). A high school student (second author) was involved in the design and implementation of the games to set the appropriate level of explanation.

*Game design*

Eight games were designed and are listed in Table 3. The games are divided into four modules with two games each: one Beginner game and one Advanced game. Students are expected to complete the Beginner game before moving onto its optional Advanced counterpart. This variable path through the eight games is shown in Figure 1(A). Topics including basic imaging parameters, imaging anatomy and flow, MR physics, and projections were covered to build a bridge from high school curricula to the imaging process. All games operate in 'virtual mode" where the image acquisition part is performed through Virtual Scanner simulation. In addition, the games were designed for communicating with an educational scanner (Figure 1(B)).

A traditional laboratory manual was developed for each game and could be used in classroom settings with didactic emphasis (see Additional Files). The game-interfaces themselves contain feedback and instructions and could be used alone in more flexible settings. A reward system was incorporated that asks the player to answer multiple choice questions and perform a series of tasks. One to five stars were awarded to the player for answering enough questions correctly and/or performing enough tasks successfully. Table 4 shows how the taught concepts can be matched to coursework at the High School, Undergraduate, and Graduate levels. The following sections describe each module in detail.

*Module 1: Imaging Basics*

The first module introduces the two basic spaces of MRI data: image space and k-space. Game 1, "What's in an image?", explains fundamental terminology in digital imaging, such as Field-of-View (FOV), matrix size, resolution, and windowing. Digital images are simulated using these different parameters. After familiarizing themselves with the settings, players are asked to change settings to generate target images under constraints and are awarded stars for choosing appropriate parameters. Following these image explorations, Game 2, "K-space magik", introduces the concept of k-space, a key mathematical representation of MR images and where data is directly collected in MR experiments. For each object imaged, the k-space can be manipulated with subsampling and erasing tools. This is then transformed back to the image space to visualize its effects which often appear as image artifacts. The interface provides an array of different signals and images to choose from and lets students draw their own image (or signal) to be converted to k-space (or spectrum) and manipulated with the same tools. Players are guided step-by-step through instructions and creative exercises to develop their intuition about k-space.

*Module 2: Imaging Biology*

The second module connects MRI to biological parts and processes. Game 3, "Brains, please", demonstrates how MR parameters affect contrast on the brain anatomy. The interface uses the MR signal equation to show how contrast, defined as signal differences among tissue types, changes with imaging parameters such as repetition time, echo time, and flip angle. A Brainweb discrete-tissue model was used (Collins et al., 1998). Players are asked to generate specified contrasts by selecting parameters and are awarded stars when the goals are achieved. In contrast to the static nature of anatomical scans, Game 4, "Fresh Blood", shows that MRI can also visualize dynamic physiological processes like blood flow. Two types of angiographic images ("bright blood" and "dark blood") can be acquired on a water flow phantom. For each sequence type, an interactive visualization panel demonstrates the principles of generating flow-based contrast, and an acquisition panel allows creation of flow rate contrast images using chosen parameters. The students can explore how each sequence parameter or the flow rate affects the contrast.

*Module 3: Imaging physics*

Module 3 focuses on the basic geometrical and physical principles of MRI. Game 5, called "Spin's got moves", allows students to see the four moves of the sample magnetization vector, the source of all MR signals. Three ways of interacting with the magnetization are provided: the main magnetic field, the Radiofrequency (RF) pulse, and the signal reception coil. Concepts of equilibrium magnetization, variable frequency precession, the rotating frame, and signal reception geometry were incorporated. Virtual Scanner Bloch simulation functions were employed for generating the magnetization dynamics (Tong et al., 2019). The students are first guided to explore the different moves of the magnetization. Then, they are asked to move a given initial magnetization to a designated place using one or more RF pulses. To take it to the next step, Game 6, "Fresh Blood", explores the $T_1$ and $T_2$ effects which describe how long the magnetization takes to return to equilibrium, affecting signal strength and contrast between tissues. Intuition is developed on two relaxation processes with signal curves and 3D vector animation. Effects of the two time constants on magnetization development are further shown by applying 90-degree and 180-degree RF pulses at various timings. Subsequently, the fundamental concept of quantitative imaging is demonstrated with $T_1$ and $T_2$ mapping. The student may choose a Region-of-Interest (ROI) from the imaged object and perform curve fitting to find the correct $T_1$ or $T_2$ value. They can also perform a full mapping process of the object to obtain $T_1$ and $T_2$ maps.

*Module 4: Imaging by Projection*

Finally, Module 4 deals with the concept of projections, useful for understanding multiple medical imaging modalities including MR and Computed Tomography (CT). Game 7, "Puzzled by Projection I", introduces students to projection imaging. Pre-generated 3D models can be pulled from the database and projected into images: 3D to 2D projection imaging at orthogonal axes and 2D to 1D projection imaging at arbitrary angles. After a few guided trials, the students are asked to predict projections resulting from a random 3D/2D model and a given projection angle. They are awarded points for making correct predictions. Game 8, "Puzzled by Projection II", turns Game 7 around and asks the student to guess the shape of an unknown 3D or 2D object by performing a limited number of projections. There are three panels: one for selecting 2D/3D mode, loading the question including the correct mystery object and three other wrong choices, and acquiring a projection; one for showing the already acquired projections (up to two for 3D and up to five for 2D); the last one for displaying the options and checking answers.

*Feedback collection*

A Google Form was created to collect feedback from game players that includes questions about previously taken science and mathematics courses, level of difficulty, and suggestions (see Additional Files). This form was used for both beta testing and the pilot deployment. Beta testing consisted of three doctoral students, one postdoctoral researcher, and one faculty member working in the field. Following beta testing, we addressed the feedback in the 1.0.0 release. The games were piloted among five high school to early college level students interested in MR over a month of exploration through a link to the deployed web application.

**Results**

*Software release*

The entire code base is available at an open-source repository hosted on Github with the GNU General Public License v3.0 (imr-framework, 2022/2022). Local installation can be achieved by users familiar with the Python programming language and its associated virtual environment setups. To enable wider distribution, the application was deployed on Heroku which renders it accessible through a web link (*Virtual Scanner Tabletop*, n.d.).

*Game structure*

The games are unified in their structure, which has three parts: (i) user input fields; (ii) visualizations; (iii) the laboratory manual. The user inputs let students interact with the experiment: for example, in Game 6: Relaxation Station, they can change the array of inversion times when performing $T_1$ mapping. Some inputs provide more freedom: for example, the canvas in Game 2 allows 2D images and 1D curves to be drawn or erased with a mouse and is used for exploring the properties of Fourier transforms. These inputs are put through the experimental "black box" which simulates subprocesses in imaging. On the other side of the "black box", visualizations complete the feedback loop as the user observes changes caused by their inputs and adjusts them to produce desired outcomes. Lastly, the lab manuals provide assistance with three components (Figure 2): first, the Instructions panel outlines one clear path for experiencing all aspects of each game: each game consists of three to five main steps, each in turn made up of smaller substeps that guide the user through the interface; second, the Definitions panel explains the terms and concepts in detail for those interested; third, the Quiz panel tests the students' understanding through multiple choice questions.

The webpage allows the user to make their own attempts at finding inputs that generate useful visualizations. This trial-and-error process was designed to help them grasp the underlying imaging subprocess. More information can be found inside the interfaces themselves, and Figure 3 gives an overview of Game 2 (advanced) and Game 5 (beginner), which have functionalities that can also be useful in regular university or graduate level MR class settings. Screenshots of all games and traditional laboratory manuals are included in Additional Files.

*Feedback*

Game-wise survey results are shown in Figure 4 for the beta testers and students. We note that there was significant layout change and optimization between beta testing and deployment based on the feedback from beta testing. Specifically, we added the sidebar containing instructions, definitions, and questions, replaced form submission with sockets to improve the parameter setting process, and unified layout across games. Among beta testers, the average score for recommendation was 4.8/5 for high school students and college students, 4.6/5 for science/engineering graduate students, and 5/5 for MR educators. Ease of use based on game interface instructions was 3.8/5 and the level of reward was 4/5. In deployment, the average recommendation scores were 3.8/5 for high school students, 4.2/5 for college students, 4/5 for science/engineering graduate students, and 4.8/5 for MR educators. Ease of use was 3.4/5 and the level of reward was 4.2/5.

**Discussion**

*Feedback and usage cases*

We designed Virtual Games to focus on illustrating concepts with interactive visualizations that allow free experimentation. This is in contrast to existing software that either conform with clinical console interfaces or have more sophisticated simulation capabilities. Beta testing feedback was used to improve user experience by unifying the layout and including the lab manual for all games. We envision three major use cases: first, students may take a complete journey through the four beginner games or all eight games over one to two months under teacher guidance; second, MR instructors may use specific visualizations to illustrate a point, such as k-space undersampling, the phase relationship between RF pulses and transverse magnetization, and quantitative mapping steps; third, students may find it helpful to access individual games on their own time to conduct a longer exploration or even contribute to new games and functionalities as a research project.

Our current implementation has limitations: feedback showed that while the updates after the beta testing reduced navigation difficulty for some games, there is still room for improvement. First, there are missing pieces to the MR pipeline from physics to images: to name a few, the idea of sequence diagrams, gradients, and the physics of precession were not included in order to keep the level accessible. Second, aspects of concept visualization and the tasks structure were more highly rated than navigation and didactic components including explanations and questions. Thus, for teaching a complete MR course, the games would work best when included into an existing curriculum. Future work will focus on expanding and revising game functionalities such as interface feedback and hints, simulation capabilities, visualization clarity, and overall flow based on user feedback.

*Hardware connection*

We previously demonstrated the feasibility of connecting Virtual Scanner Games via the Red Pitaya 122.88 to the open-source Magnetic Resonance Control System (MaRCoS) to generate and visualize pulse sequence waveforms on an oscilloscope (Negnevitsky et al., 2022; Tong et al., 2023). MaRCoS is compatible with peer-reviewed and published open-source hardware (Cooley et al., 2020). While the experiment shows promise for the hardware mode, more work is needed to enable it across the games and ensure consistent data can be collected under different settings: for example, both in the lab and in a high school classroom. Mechanisms for switching and indicating the current mode, auto-calibration of the central frequency and RF power, and instructions accounting for errors must be incorporated. Other educational possibilities including custom design and execution of pulse sequences will be explored in the future.

*Local deployment mode and accessible MR education*

To make the educational materials accessible to the world, efficient local deployment and usage where there is no fast internet connection, or where internet connection is geographically limited, is essential. Currently, local usage of Virtual Scanner Games is possible by installing through the PyPI index and copies of the codebase may be made and distributed with physical flash drives. This procedure can pose problems with Python versions and library dependencies and should be further optimized. On the other hand, where there is central access to internet access such as in high schools and universities, students can engage in a flexible and shared manner without the need for installation. We plan to partner with high schools through existing extracurricular educational programs to distribute the games.

**Conclusion**

We designed, implemented, and deployed eight educational games about MRI basics on a web interface. All components are open-source. Virtual Scanner Games is, to our knowledge, the first of its kind in MR education as a zero-footprint browser application that is extendable for operating open-source MR hardware. Multiple educational scenarios including classroom, personal, and hybrid were considered in the design and implementation.

**List of Abbreviations**

**$B_0$:** main magnetic field

**FID:** Free Induction Decay

**FOV**: Field of view

**GUI:** Graphical user interface

**IRSE:** Inversion Recovery Spin echo

**MaRCoS:** Magnetic Resonance Control System

**MR:** Magnetic Resonance

**MRI**: Magnetic resonance imaging

**NMR:** Nuclear Magnetic Resonance

**RF:** Radiofrequency

**ROI:** Region of interest

**SE:** Spin Echo

**STEAM:** Science, Technology, Engineering, Art, and Mathematics

**$T_1$:** Longitudinal relaxation time

**$T_2$:** Transverse relaxation time

**List of additional files**

Survey Questions (1): feedback_form.pdf

Screenshots of all games (1): game_screenshots.docx

Lab manuals (8): game1.docx, game2.docx, game3.docx, game4.docx, game5.docx, game6.docx, game7.docx, game8.docx

https://www.ismrm.org/2019-junior-fellow-challenge/africa/

Anazodo, U. C., Ng, J. J., Ehiogu, B., Obungoloch, J., Fatade, A., Mutsaerts, H. J. M. M., Secca, M. F., Diop, M., Opadele, A., Alexander, D. C., Dada, M. O., Ogbole, G., Nunes, R., Figueiredo, P., Figini, M., Aribisala, B., Awojoyogbe, B. O., Aduluwa, H., Sprenger, C., … the Consortium for Advancement of MRI Education and Research in Africa (CAMERA). (2023). A framework for advancing sustainable magnetic resonance imaging access in Africa. *NMR in Biomedicine*, *36*(3), e4846. https://doi.org/10.1002/nbm.4846

Bakhmutov, V. I. (2011). *Solid-State NMR in Materials Science: Principles and Applications*. CRC Press.

Benoit-Cattin, H., Collewet, G., Belaroussi, B., Saint-Jalmes, H., & Odet, C. (2005). The SIMRI project: A versatile and interactive MRI simulator. *Journal of Magnetic Resonance*, *173*(1), 97–115. https://doi.org/10.1016/j.jmr.2004.09.027

Collins, D. L., Zijdenbos, A. P., Kollokian, V., Sled, J. G., Kabani, N. J., Holmes, C. J., & Evans, A. C. (1998). Design and construction of a realistic digital brain phantom. *IEEE Transactions on Medical Imaging*, *17*(3), 463–468. https://doi.org/10.1109/42.712135

Cooley, C. Z., Stockmann, J. P., Witzel, T., LaPierre, C., Mareyam, A., Jia, F., Zaitsev, M., Wenhui, Y., Zheng, W., Stang, P., Scott, G., Adalsteinsson, E., White, J. K., & Wald, L. L. (2020). Design and implementation of a low-cost, tabletop MRI scanner for education and research prototyping. *Journal of Magnetic Resonance*, *310*, 106625. https://doi.org/10.1016/j.jmr.2019.106625

Geethanath, S., & Vaughan Jr., J. T. (2019). Accessible magnetic resonance imaging: A review. *Journal of Magnetic Resonance Imaging*, *49*(7), e65–e77. https://doi.org/10.1002/jmri.26638

Hackländer, T., & Mertens, H. (2005). Virtual MRI: A PC-based simulation of a clinical MR scanner1. *Academic Radiology*, *12*(1), 85–96. https://doi.org/10.1016/j.acra.2004.09.011

imr-framework. (2022). *Virtual Scanner Tabletop Web Games* [JavaScript].
14

**Figures**

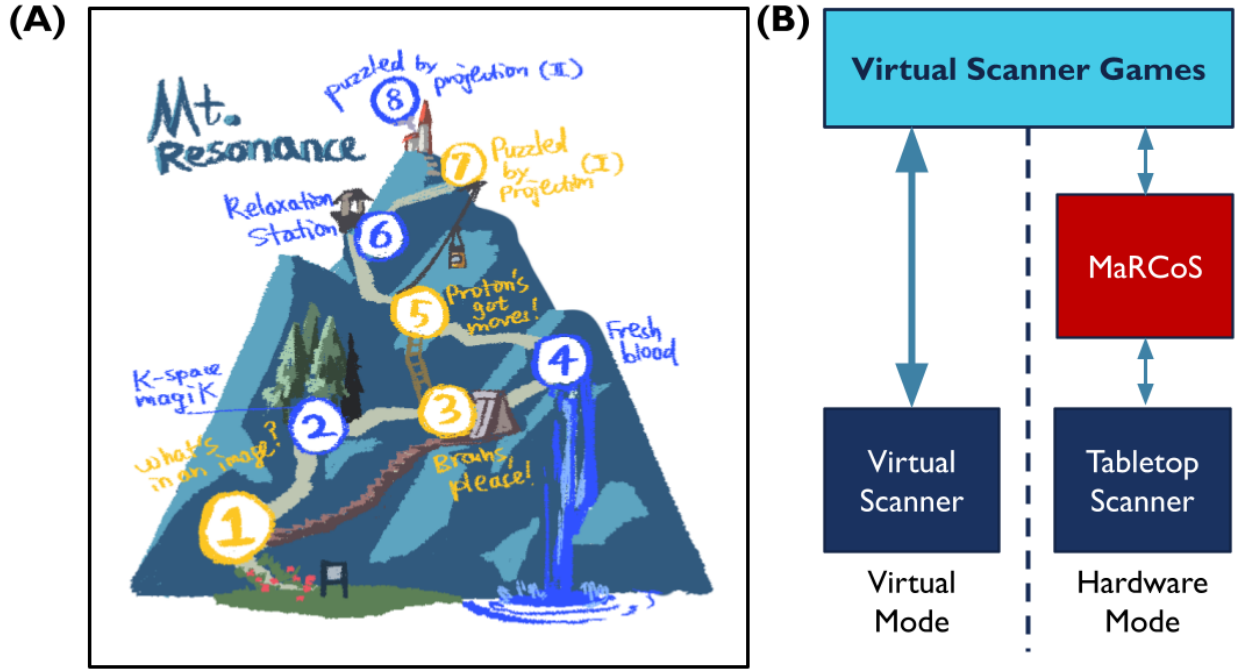

Figure 1: Conceptual diagrams of Virtual Scanner Games. (A) The games are represented as a journey up Mr. Resonance with eight stops along the way. The beginner games (1,3,5,7) make up the main route while the advanced games (2,4,6,8) are optional side stops for more enthusiastic or experienced students; (B) Connection of Virtual Scanner Games with existing software. In virtual mode, images are simulated either with game-specific code or with useful classes from the Virtual Scanner library. In hardware mode, a tabletop instructional scanner would be controlled through the MAgnetic Resonance COntrol System (MaRCoS) (Negnevitsky et al., 2022).

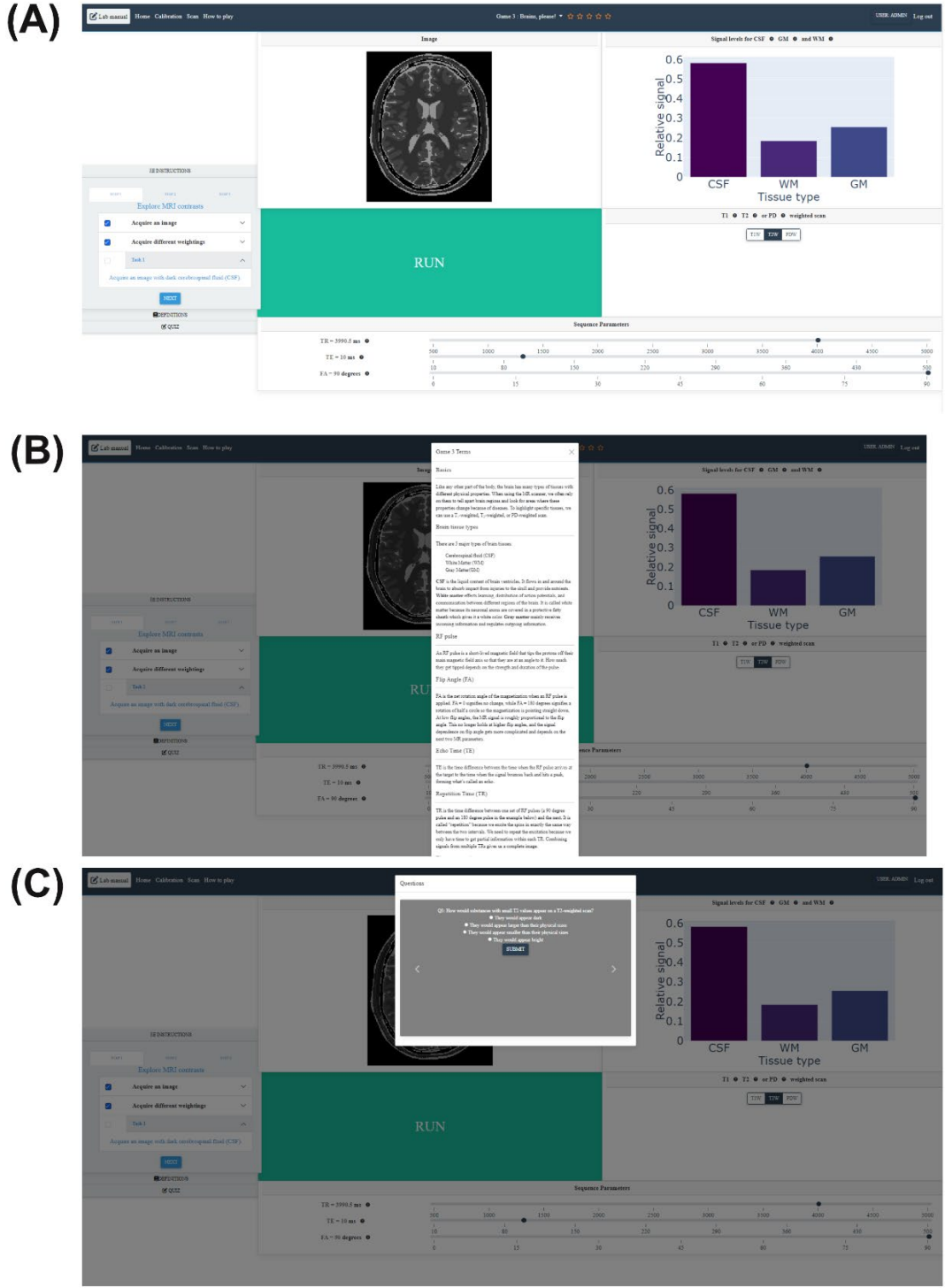

Figure 2: Lab manual components inside the interface for Game 3. The layout is shared by all games. (A) Regular Game 3 interface with the sidebar opened and showing the instructions panel; (B) Game 3 with the definitions panel open; (C) Game 3 with the quiz panel open.

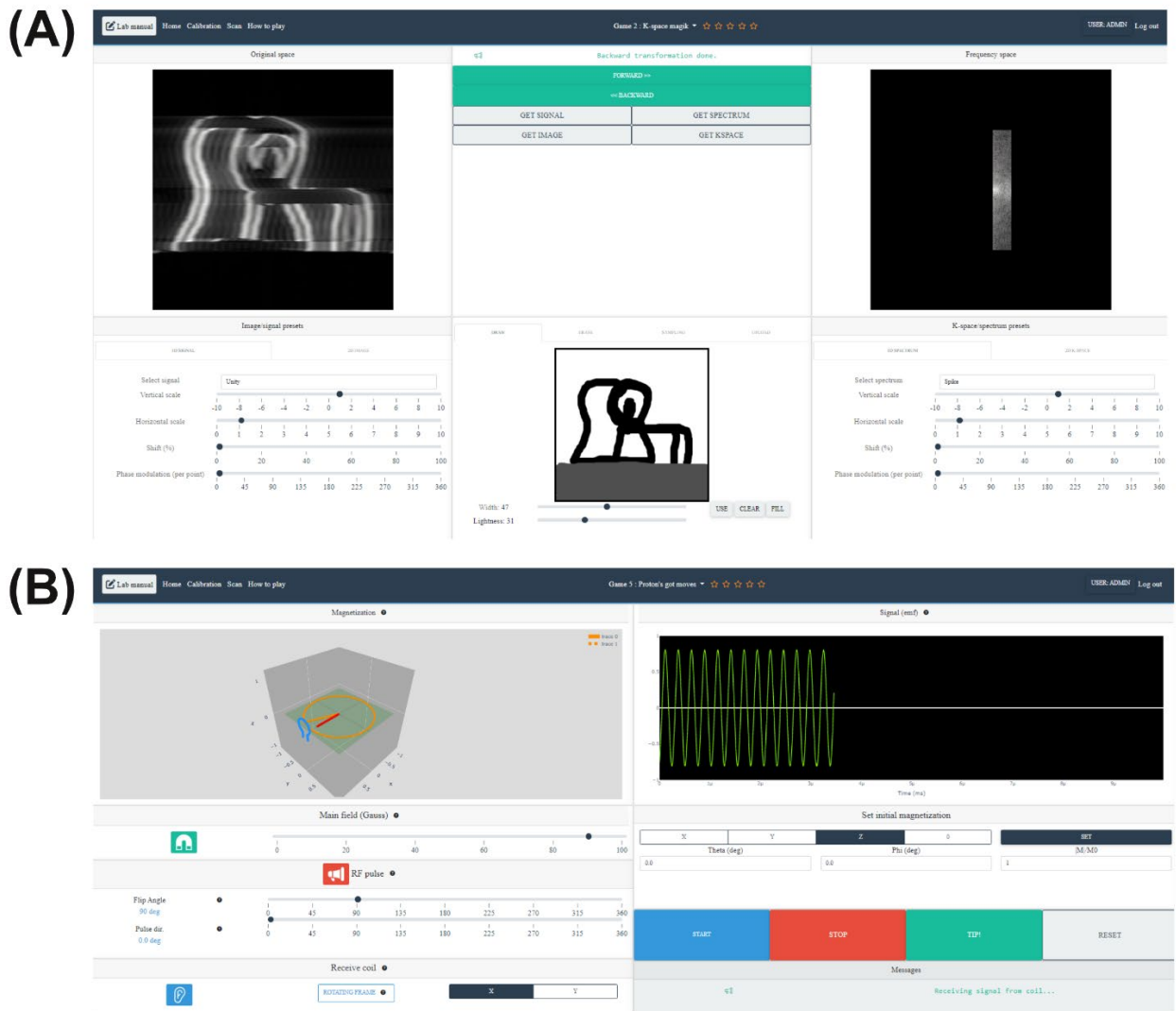

Figure 3: Example games with features useful in MRI classes. (A) Game 2: K-space magik allows generation and manipulation of k-space (deleting, skipping lines) which is reflected on the image; (B) Game 5: Proton's got moves simulates magnetization dynamics under the Bloch equation with no relaxation terms, which helps explain concepts such as precession, the rotating frame of reference, RF nutation, and signal induction.

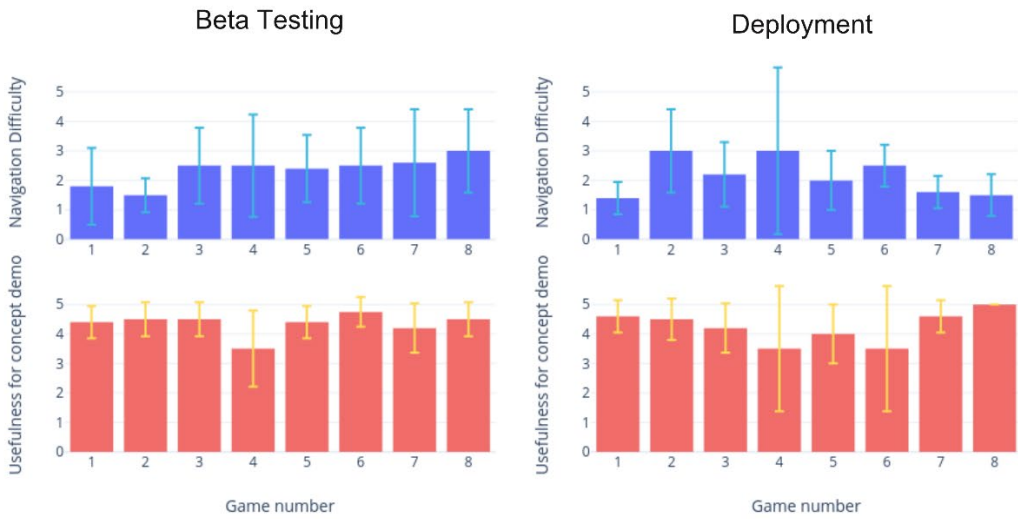

Figure 4: Averaged survey results from beta testing and deployment. The scale for each question is from 1 to 5 and error bars show standard deviation. For beta testing, all beginner games (1,3,5,7) had five players and all advanced games (2,4,6,8) had four; for deployment, all beginner games had five players and all advanced games had two. We note that the second author who mainly contributed to two games in the beta testing version of the software was included in the deployment survey after playing the 1.0.0 release version of all eight games.

# Tables

| Name | Open source release | Language | GUI type | Target trainees | Features |
|---|---|---|---|---|---|
| SIMRI (Benoit-Cattin et al., 2005) | No | C | Local | Master's students | Imaging based on Bloch simulation with $T_2^*$ and off-resonance; 1D custom experiment interface |
| Virtual MRI (Hackländer & Mertens, 2005) | Yes | Java | Local | Medical students and interns | Imaging based on analytical signal expression, artifacts, k-space manipulations, motion |
| JEMRIS (Stöcker et al., 2010) | Yes | MATLAB, C++ | Local | Nonspecific | Imaging based on Bloch simulation with sequence design functions; includes $T_2^*$ and off-resonance, movement, flow effects |
| MRiLab (F. Liu et al., 2017) | Yes | MATLAB, C++ | Local | Nonspecific | Imaging based on Bloch simulation with exchange model, including motion, hardware maps, and arbitrary pulse sequence |
| A virtual scanner for teaching fundamental MR (Wilhjelm et al., 2018) | No | MATLAB | Local | Engineering students studying Physics of Medical Imaging | Free Induction Decay (FID) signal, RF pulse, demodulation, and image visualizations |
| MR SPRINT (Pizetta, 2018) | No | Python, C++ | Local | Nonspecific | Phantom, $B_0$ map, and Bloch simulation of FID |

| Name / citation | | Open source | Language | Platform | Target audience | Description |
|---|---|---|---|---|---|---|
| Virtual Scanner (Tong et al., 2019) | | Yes | Python | Web | Radiographers | Imaging based on Bloch simulation of arbitrary Pulseq files with pre-defined phantoms |
| Web-based educational MR simulator (Treceño-Fernández et al., 2019) | | No | Python, C++ | Web | Radiographers | Imaging based on analytical signal expression, geometrical planning, artifacts and noise, $\Delta B_0$ |
| (This work) Virtual Scanner Games | | Yes | Python | Web | High school students and beyond | Interactive tutorials on fundamental aspects of MRI (digital imaging, biology, physics, and spatial thinking) |

Table 1: Open-source MR simulation software with educational goals or potential compared to the proposed solution (last row)

| Name / citation | Type | Magnet | Field strength | Imaging | Sample volume diameter | Cost | Weight | Console software | Teaching Materials |
|---|---|---|---|---|---|---|---|---|---|
| Desktop MR imaging system (Wright et al., 2001) | Research | Permanent | 0.21T | Yes | 20 mm | $13,500 | N/A | LabVIEW | Not provided |
| Mobile NMR relaxometry platform. (Twieg et al., 2013) | Open hardware | Surface magnet | 0.196T | No | N/A | $750 | 1.3 kg | MATLAB | Not provided |
| Magritek Spinsolve | Commercial | Permanent | ~1T | No | 5 mm | $43,725 | 55 kg | Included | Workbook |
| Niumag EDUMR (Niumag, n.d.) | Commercial | Permanent | 0.5T | Yes | 15 mm | $32,000 | 138 kg | Included | Workbook |
| Pure devices | Commercial | Permanent | 0.43T | Yes | 10 mm | $43,000 | 15.8 kg | Included | Software |

| benchtop MRI (Pure Devices, n.d.) | | | | | | | | | |
|---|---|---|---|---|---|---|---|---|---|
| Martinos Tabletop Instructional Scanner (Cooley et al., 2020) | Open hardware | Permanent | 0.18T / 0.37T | Yes | 10 mm | $10,000 | 13 kg (magnet) | MATLAB | Lab y/manuals and course notes |

Table 2: Examples of educational MR hardware solutions. The list was compiled with the authors' best knowledge from English-language online resources but may not be comprehensive in representing all similar systems around the world.

| Module | # | Name | Concepts |
|---|---|---|---|
| **Basic Imaging** | 1 | What's in an image? | FOV, resolution, windowing |
| | 2 | K-space magiK | spectra, k-space, sampling |
| **Imaging Biology** | 3 | Brains, please | Brain tissue and sequence parameters |
| | 4 | Fresh blood | Flow imaging |
| **MR Physics** | 5 | Proton's got moves | Precession, nutation, signal reception |
| | 6 | Relaxation station | $T_1$ and $T_2$ relaxation |
| **3D** | 7 | Puzzled by Projection I | 1D and 2D Projections |

| Module | # | Name | Concepts |
|---|---|---|---|
| **Basic Imaging** | 1 | What's in an image? | FOV, resolution, windowing |
| | 2 | K-space magiK | spectra, k-space, sampling |
| **Imaging Biology** | 3 | Brains, please | Brain tissue and sequence parameters |
| | 4 | Fresh blood | Flow imaging |
| **Thinking** | | | (forward) |
| | 8 | Puzzled by Projection II | 1D and 2D Projections (inverse) |

Table 3: The eight Virtual Scanner Games

| | Concepts | Games | High School | Undergraduate | Graduate |
|---|---|---|---|---|---|
| **MR expertise level** | N/A | N/A | Limited experience in imaging | Fundamental math/engineering background and some | Advanced engineering/math background and some |

|  |  |  |  | understanding of biomedical imaging | understanding of MR physics |
|---|---|---|---|---|---|
| **Learning goals** | N/A | N/A | Gain intuition on what MRI does and how | Understanding physical principles of MRI | Design, acquire, reconstruct, analyze |
| **Science** | Magnetization dynamics, brain anatomy | 3,5,6 | AP Physics, Neuroscience, AP Biology | Electromagnetism, Introduction to Neuroscience | Electrodynamics |
| **Technology** | 3D printing, MR hardware | 2,4,7,8 | Electronics, 3D modeling | Electronics, 3D modeling and printing | Electronics, 3D modeling and printing |
| **Engineering** | Sequence parameters | 1,2,3,4,7,8 | AP Calculus AB/BC, AP Physics | Biomedical Imaging, Magnetic Resonance Imaging | Biomedical Imaging, Magnetic Resonance Imaging |
| **Art** | Digital imaging | 1,7,8 | Visual Arts | Digital image processing | Digital image processing |

| Mathematics | Fourier transform | 1,2,4,5,6,7,8 | AP Calculus AB/BC | Signal processing | Signal processing |

Table 4: Connecting games to school curricula at different levels.

**Declarations**

*Availability of data and materials:*

- Project name: Virtual Scanner Games
- Project home page: https://github.com/imr-framework/vs-tabletop
- Operating system(s): Platform independent
- Programming language: Python, Javascript
- Other requirements: Python 3.9
- License: GNU General Public License v3.0
- Any restrictions to use by non-academics: None

*Competing interests:* The authors declare that they have no competing interests.

*Funding:* This work was supported by the Center for Engineering and Precision Medicine's joint research award (PI: Geethanath), Friedman Research Scholar award, Friedman Brain Institute (PI: Geethanath and Taufique), NIH 1U01 EB025153-01 grant and performed at the Accessible MRI Laboratory, Icahn School of Medicine, Zuckerman Mind Brain Behavior Institute, and the Columbia MR Research Center.

*Authors' contributions:*

GT designed and coded the Virtual Scanner games. She authored the manuscript. RA assisted in developing concepts for all games, co-implemented Games 1 and 2, and created feedback forms. JTV helped revise the manuscript. SG conceptualized the virtual scanner project as a standalone MRI system level simulator and web-based games as an MR dissemination tool for students in Africa in particular and high school students in general, supervised the project, contributed to design features, gathered volunteers for testing, and helped revise the manuscript. All authors read and approved the final manuscript.

*Acknowledgements:* We thank Kunal Aggarwal, Tiago Fernandes, Marina Manso Jimeno, and Enlin Qian for their beta-testing and Grace Kuriakose, Nafisa Promi, Oiya Ivan Etoku, and Nishika Girish for going through the games and providing valuable feedback. We were also grateful for the help and support on the hardware connection from Leeor Alon, Akbar Alipour, Lawrence Wald, Jason Stockmann, Benjamin Menküc, and Vlad Negnevitsky.

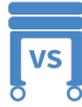

Landing page screenshot

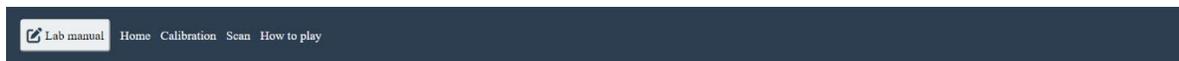

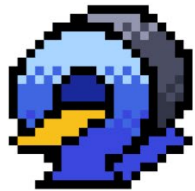

Login page screenshot

Game selection screenshot

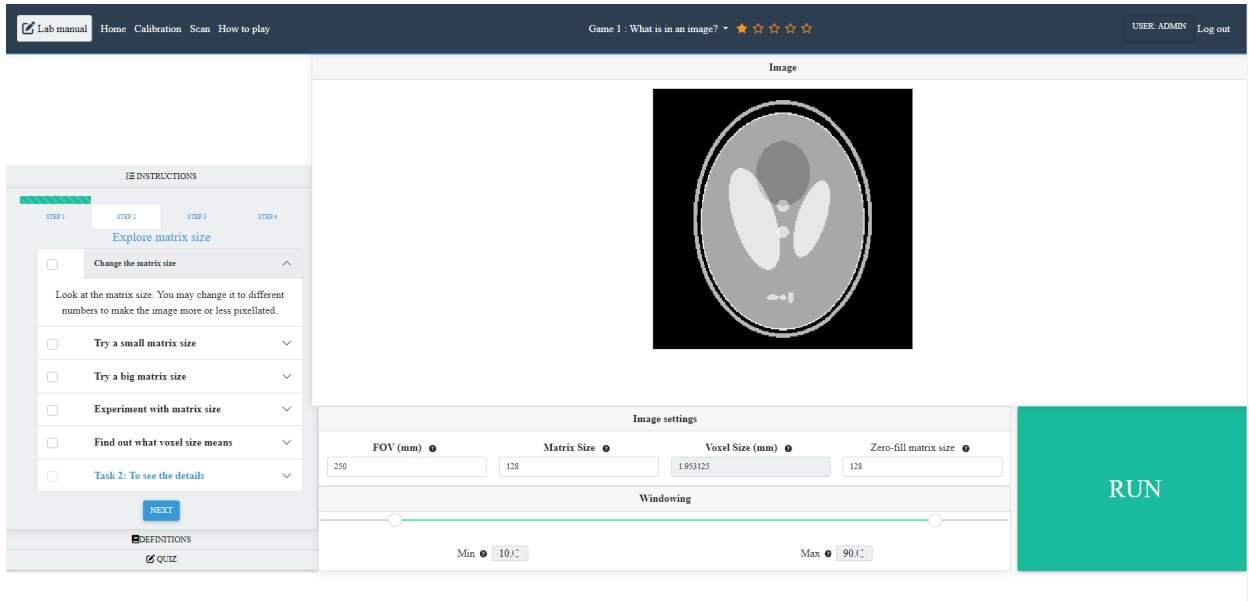

Game 1 screenshot

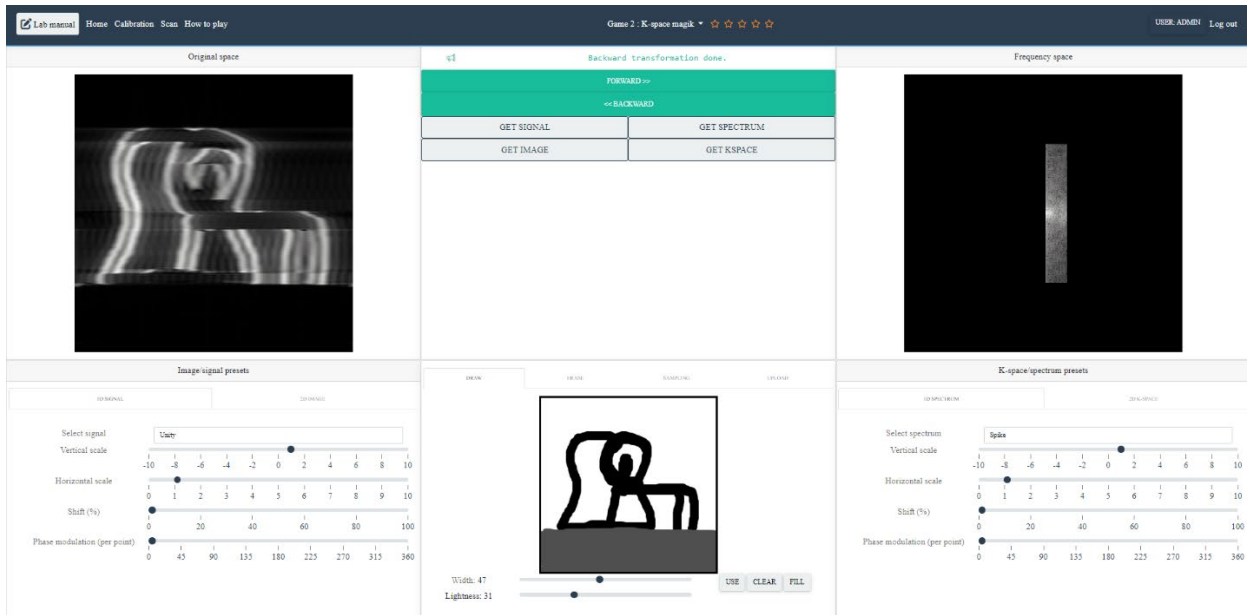

Game 2 screenshot

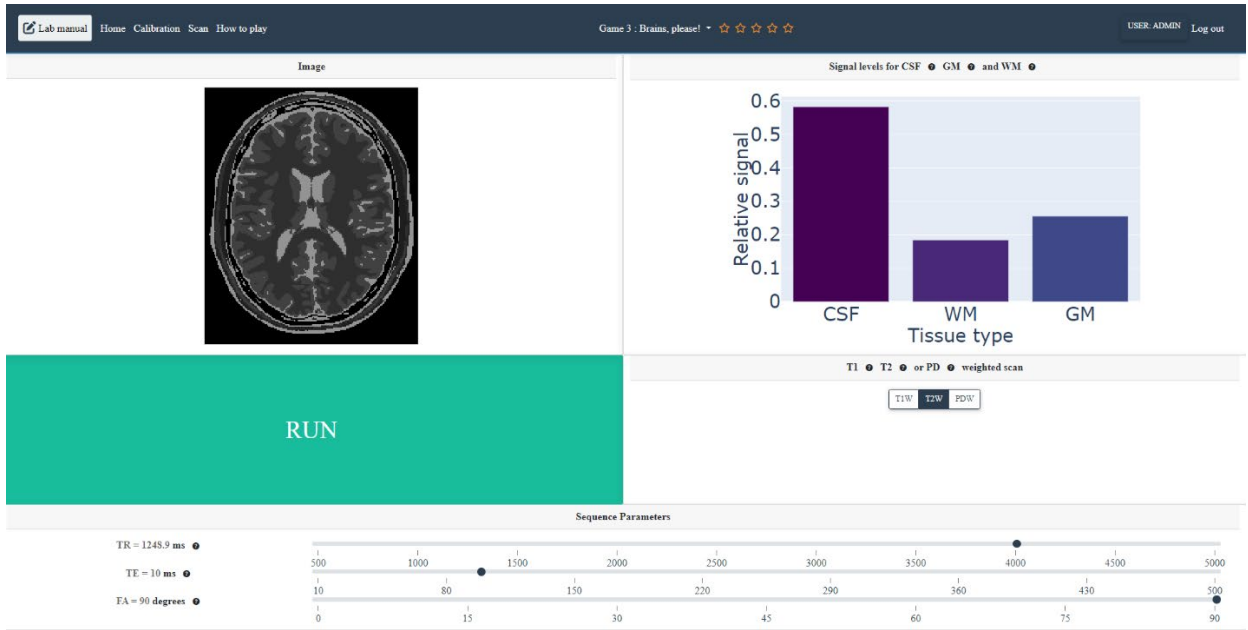

Game 3 screenshot

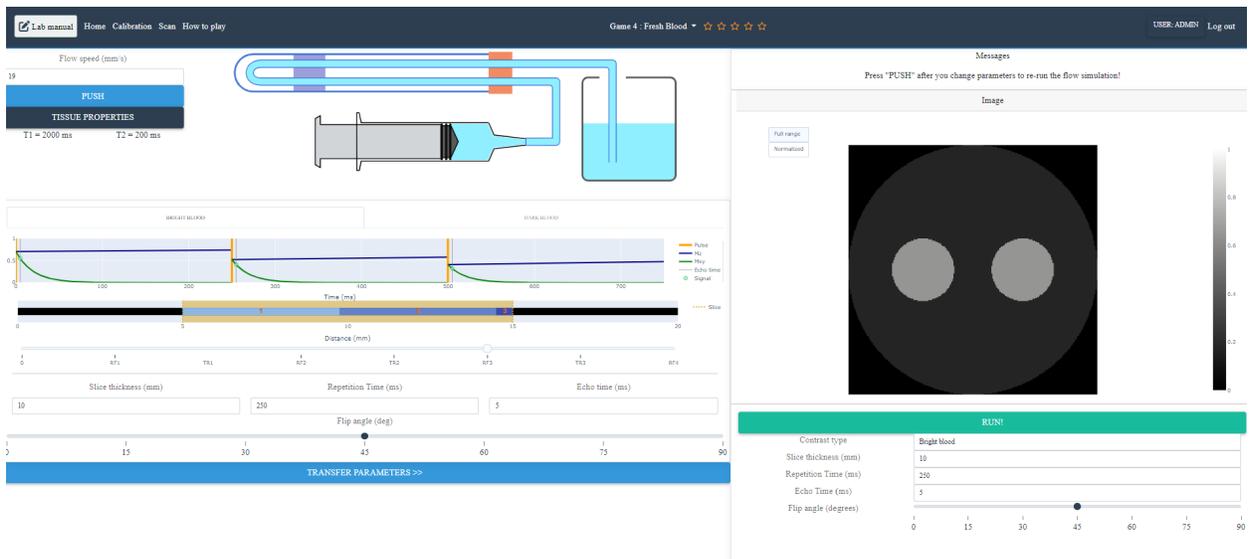

Game 4 screenshot

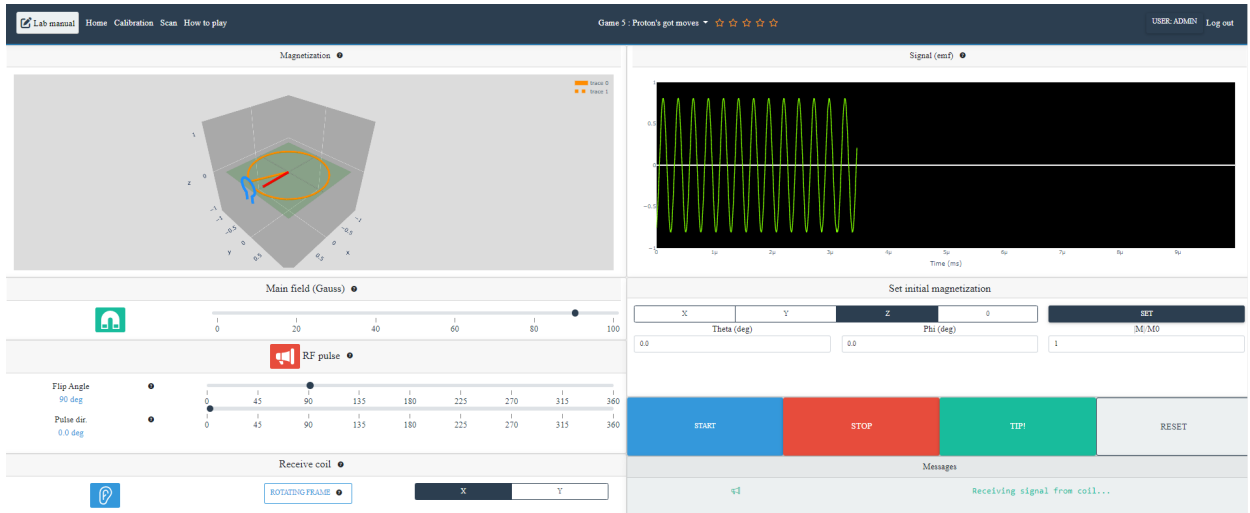

Game 5 screenshot

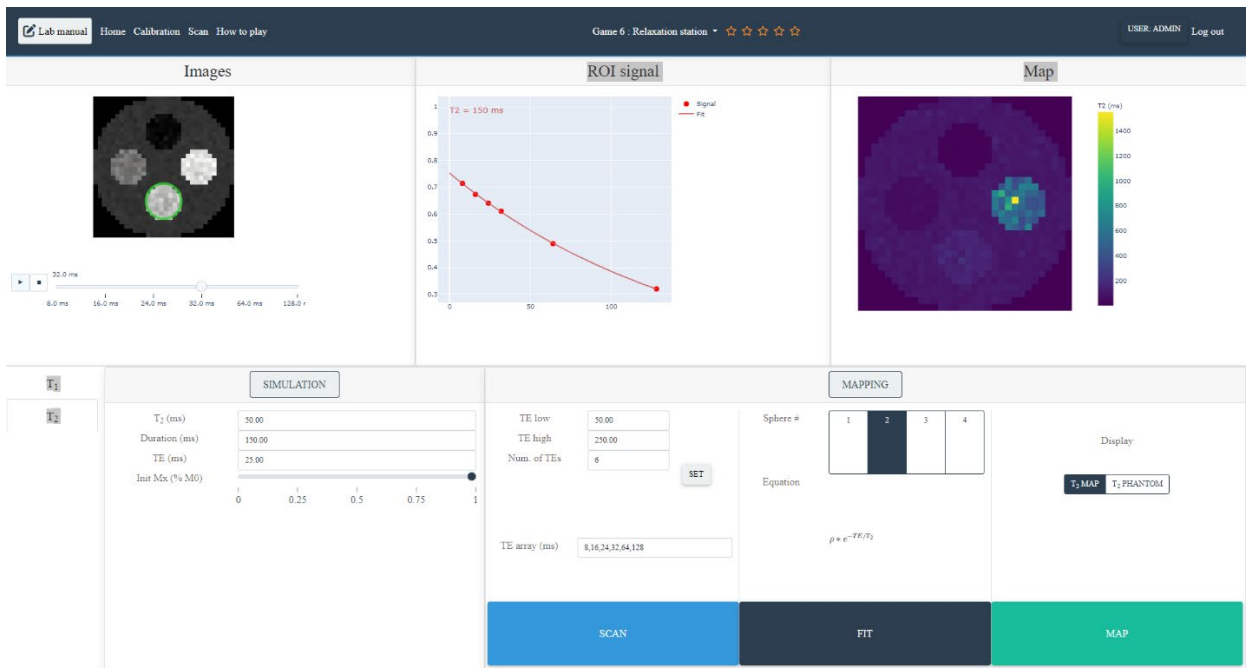

Game 6 screenshot

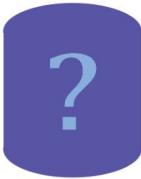
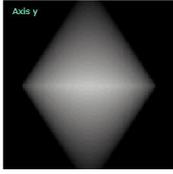
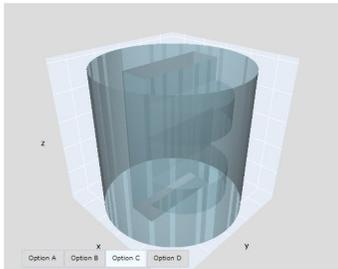

Game 7 screenshot

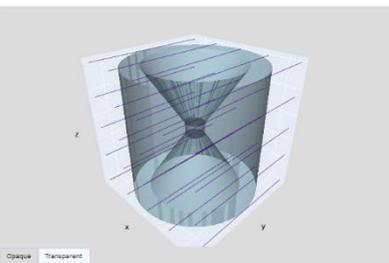

Game 8 screenshot

# VS Tabletop Games Feedback Form

* Required

1. Name (email) *

2. Role *

   *Mark only one oval.*

   - ◯ High school student / graduate
   - ◯ College student / graduate
   - ◯ Masters student / graduate
   - ◯ PhD student
   - ◯ PhD graduate / Post-doc
   - ◯ Faculty
   - ◯ Other: _______________

3. What games have you played? *

   *Check all that apply.*

   - ☐ Game 1: What is in an image?
   - ☐ Game 2: K-Space Magik
   - ☐ Game 3: Brains, Please
   - ☐ Game 4: Fresh Blood
   - ☐ Game 5: Protons got moves
   - ☐ Game 6: Relaxation Station
   - ☐ Game 7: Puzzled by projection 1
   - ☐ Game 8: Puzzled by projection 2

## Science courses experience

4. What science course did you enjoy the most? *

   Check all that apply.

   - [ ] Physics
   - [ ] AP Physics 1
   - [ ] AP Physics 2
   - [ ] AP Physics C
   - [ ] Biology
   - [ ] AP Biology
   - [ ] Chemistry
   - [ ] AP Chemistry

5. Which of the following math courses have you completed? *

   Check all that apply.

   - [ ] Algebra 1
   - [ ] Geometry
   - [ ] Algebra 2
   - [ ] Pre-Calculus
   - [ ] Calculus AB/1
   - [ ] Calculus BC/2
   - [ ] Other: _______________

## Game-wise feedback

If you haven't played a game, you may skip questions related to it.

6. How did you feel about navigating Game 1?

    *Mark only one oval.*

    Very Easy

    1 ○
    2 ○
    3 ○
    4 ○
    5 ○

    Very Difficult

7. How useful do you find Game 1 for understanding and demonstrating the relevant concepts?

    *Mark only one oval.*

    Not useful

    1. ○
    2. ○
    3. ○
    4. ○
    5. ○

    Very useful

8. What do you think is the hardest concept(s) to comprehend in Game 1?

    *Check all that apply.*

    ☐ Field-of-View (FOV)
    ☐ Matrix Size
    ☐ Voxel Size
    ☐ Windowing
    ☐ Zero fill
    ☐ Other: ______________

9. How did you feel about navigating Game 2?

   *Mark only one oval.*

   Very Easy

   1 ○
   2 ○
   3 ○
   4 ○
   5 ○

   Very Difficult

10. How useful do you find Game 2 for understanding and demonstrating the relevant concepts?

    *Mark only one oval.*

    Not useful

    1. ◯
    2. ◯
    3. ◯
    4. ◯
    5. ◯

    Very useful

11. What do you think are the hardest concept(s) to comprehend in Game 2?

    *Check all that apply.*

    ☐ Signal and spectrum
    ☐ Spectrum
    ☐ K-space
    ☐ Forward transform
    ☐ Backward transform
    ☐ Other: ___________

12. How did you feel about navigating Game 3?

    *Mark only one oval.*

    Very easy

    1 ◯

    2 ◯

    3 ◯

    4 ◯

    5 ◯

    Very difficult

13. How useful do you find Game 3 for understanding and demonstrating the relevant concepts?

    *Mark only one oval.*

    Not useful

    1. ◯
    2. ◯
    3. ◯
    4. ◯
    5. ◯

    Very useful

14. What do you think are the hardest concept(s) to comprehend in Game 3?

    *Check all that apply.*

    ☐ Repetition Time (TR)
    ☐ Echo Time (TE)
    ☐ Flip Angle (FA)
    ☐ Brain tissue types (WM, GM, CSF)
    ☐ T1 weighted imaging
    ☐ T2 weighted imaging
    ☐ PD weighted imaging
    ☐ Other: _______________

15. How did you feel about navigating Game 4?

    Mark only one oval.

    Very easy

    1 ◯

    2 ◯

    3 ◯

    4 ◯

    5 ◯

    Very difficult

16. How useful do you find Game 4 for understanding and demonstrating the relevant concepts?

    *Mark only one oval.*

    Not useful

    1. ○
    2. ○
    3. ○
    4. ○
    5. ○

    Very useful

17. What do you think is the hardest concept(s) to comprehend in Game 4?

    *Check all that apply.*

    ☐ Flow contrast
    ☐ Time-of-flight imaging (bright blood)
    ☐ Spin Echo imaging (dark blood)
    ☐ Steady state
    ☐ Sequence parameters

18. How did you feel about navigating Game 5?

*Mark only one oval.*

Very easy

1 ◯

2 ◯

3 ◯

4 ◯

5 ◯

Very difficult

19. How useful do you find Game 5 for understanding and demonstrating the relevant concepts?

    *Mark only one oval.*

    Not useful

    1. ◯
    2. ◯
    3. ◯
    4. ◯
    5. ◯

    Very useful

20. What do you think is the hardest concept(s) to comprehend in Game 5?

    *Check all that apply.*

    ☐ Spin
    ☐ Main Field
    ☐ RF pulse and nutation
    ☐ Receive Coil
    ☐ Net Magnetization
    ☐ Precession
    ☐ Electromotive Force
    ☐ Free Induction Decay

21. How do you feel about navigating Game 6?

*Mark only one oval.*

Very easy

1 ◯

2 ◯

3 ◯

4 ◯

5 ◯

Very difficult

22. How useful do you find Game 6 for understanding and demonstrating the relevant concepts?

    *Mark only one oval.*

    Not useful

    1. ◯
    2. ◯
    3. ◯
    4. ◯
    5. ◯

    Very useful

23. What do you think is the hardest concept(s) to comprehend in Game 6?

    *Check all that apply.*

    ☐ T1 relaxation
    ☐ T2 relaxation
    ☐ T1 sequence (IRSE)
    ☐ T2 sequence (SE)
    ☐ Inversion time (TI)
    ☐ Echo time (TE)
    ☐ T1 mapping
    ☐ T2 mapping

24. How do you feel about navigating Game 7?

   Mark only one oval.

   Very easy

   1 ◯

   2 ◯

   3 ◯

   4 ◯

   5 ◯

   Very difficult

25. How useful do you find Game 7 for understanding and demonstrating the relevant concepts?

    *Mark only one oval.*

    Not useful

    1. ◯
    2. ◯
    3. ◯
    4. ◯
    5. ◯

    Very useful

26. What do you think is the hardest concept(s) to comprehend in Game 7?

    *Check all that apply.*

    ☐ 3D model structures
    ☐ 3D to 2D projection
    ☐ 2D to 1D projection
    ☐ How the puzzles work

27. How do you feel about navigating Game 8?

    Mark only one oval.

    Very easy

    1 ◯

    2 ◯

    3 ◯

    4 ◯

    5 ◯

    Very difficult

28. How useful do you find Game 8 for understanding and demonstrating the relevant concepts?

    *Mark only one oval.*

    Not useful

    1. ○
    2. ○
    3. ○
    4. ○
    5. ○

    Very useful

29. What do you think is the hardest concept(s) to comprehend in Game 8?

    *Check all that apply.*

    ☐ 2D projections
    ☐ 1D projections
    ☐ Projection angles
    ☐ Projection axis
    ☐ Inferring objects from their projections

Overall experience

30. What part of the games did you have the most fun with? *

    *Check all that apply.*

    - [ ] Game exploration
    - [ ] Concepts
    - [ ] Lab Manual
    - [ ] Questions
    - [ ] Game tasks
    - [ ] Other: ________________

31. What aspect do you think works the best in the games? *

    *Check all that apply.*

    - [ ] Conceptual demonstration
    - [ ] Implementation of learning goals
    - [ ] Navigation of GUI element / user experience
    - [ ] Questions, tasks, and reward system
    - [ ] Other: ________________

32. What do you think is the hardest to follow along in the lab manuals? *

    *Check all that apply.*

    - [ ] Why?
    - [ ] Key Terms
    - [ ] Definitions / Explanations
    - [ ] Lab Procedures
    - [ ] Questions

33. Which aspect(s) do you think worked best in the lab manual? *

   Check all that apply.

   - [ ] Why?
   - [ ] Key Terms
   - [ ] Definitions / Explanations
   - [ ] Lab Procedures
   - [ ] Questions

34. How much would you recommend the games to high schoolers? *

   Mark only one oval.

   Will not recommend

   1. ( )
   2. ( )
   3. ( )
   4. ( )
   5. ( )

   Will recommend

35. How much would you recommend the games to college students?

Mark only one oval.

Will not recommend

1. ◯
2. ◯
3. ◯
4. ◯
5. ◯

Will recommend

36. How much would you recommend the games to science/engineering graduate students?

    Mark only one oval.

    Will not recommend

    1. ◯
    2. ◯
    3. ◯
    4. ◯
    5. ◯

    Will recommend

37. How much would you recommend the games to Magnetic Resonance (MR) educators?

*Mark only one oval.*

Will not recommend

1. ◯
2. ◯
3. ◯
4. ◯
5. ◯

Will recommend

38. How easy was it to start playing the game with only the instructions provided by the game interface? *

Mark only one oval.

Very Hard

1 ○

2 ○

3 ○

4 ○

5 ○

Very Easy

39. How rewarding was your overall experience? *

    Mark only one oval.

    Not rewarding

    1 ◯

    2 ◯

    3 ◯

    4 ◯

    5 ◯

    Very rewarding

40. Please list any improvements we can make to the GUI of the game. *

    _______________________________________________

    _______________________________________________

    _______________________________________________

    _______________________________________________

41. How do you think we can make MR concepts easier for the students to understand? *

42. Any additional comments, questions, or suggestions?

# Game 1 Lab Manual: What's in an Image?

## Why?

Understanding the components of an image is the basis to understanding an MR image and altering it to your specifications. Components like **voxels**, **field-of-view**, and **matrix size** allow the user to manipulate the **resolution** and viewing range to achieve the specific image quality.

## Materials

- Star phantom
- Two dots phantom

## Background:

1. Key Terms
    a. Pixels/Voxels
    b. Resolution
    c. Matrix size
    d. Field-of-View (FOV)
    e. Windowing
2. Basics

    There are two main types of digital images: monochrome and color. Color images can be represented with three components: Red, Green, and Blue (RGB). It is the layering of these three colors that make up the final image. Monochrome images are made up of different gray levels. MR images are examples of monochrome images: each voxel has one number, called the image magnitude or intensity, that indicates its brightness.

3. Explanations
    a. **Pixels/Voxels**: It is a metric defining **each unit in a picture/scan**. A **pixel** defines space in a scan with **two dimensions**, while the **voxel** defines a space in **three dimensions**. MR images are made up of voxels as they always have a third dimension or thickness.

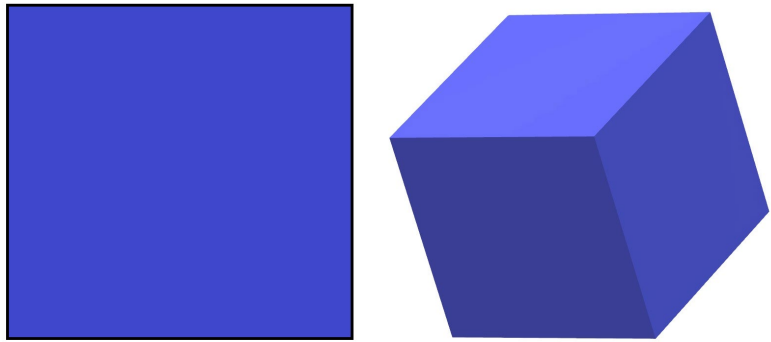

As seen in the above image, the pixel (left image) is two-dimensional and exists on the x-y plane while the voxel (right image) is three-dimensional and exists on the x, y, and z space. The pixel only has a length and a width while the voxel has a depth as well.

b. **Resolution** refers to how sharp an image is and It is related to the size of the individual pixel (2D) or voxel (3D). The smaller the pixel or voxel the greater the resolution. Voxels segment continuous space into discrete units. The signal in each voxel is an **average of all the sub-signals within it**. The smaller you make the voxels, the more closely spaced details you see because they tend to be assigned to different voxels. You **cannot see the details** when using very large voxels because the spatial details are averaged out and any locational information at smaller scales than the voxel is lost.

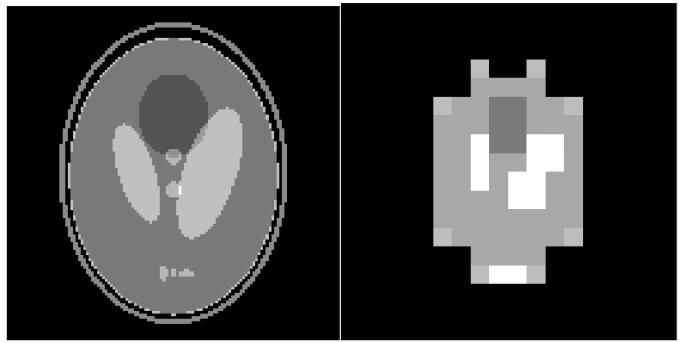

The above image contains two pictures: one with high resolution (left) and the other with low resolution (right). The higher resolution image expresses more details.

c. **Matrix size (N):** Matrix size is an integer equal to the number of voxels in each dimension. For example, an MR image can have a matrix size of 512 x 256, meaning that one side is divided into 512 parts and the other is divided into 256 parts. The image would have a total of 512 x 256 = 131072 voxels. **Small and a lot of voxels** mean the information is going to be better localized leading to **more of the details being pronounced** (remember that the more information gets averaged in each pixel, the less pronounced it will be). This is like having a high-megapixel camera.

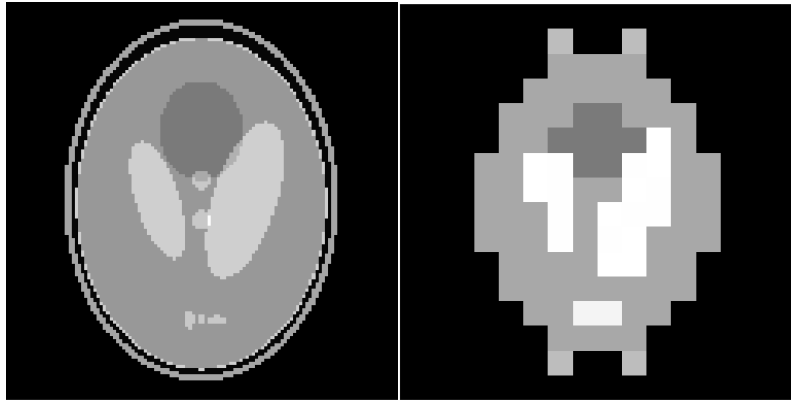

The image above has two pictures: one with many small pixels, and the other with fewer large pixels. As seen, the left picture has a higher resolution than the right one.

d. **Field-of-View(FOV):** As you zoom in on a camera, a specific part of the image is seen while the rest is discarded. Field-of-view (FOV) is usually given in millimeters in MRI and refers to how much physical distance is covered between the left and right (or top and bottom) sides of the rectangular image. When matrix size is kept the same but FOV is increased, each pixel covers a larger distance and this **lowers the image resolution**.

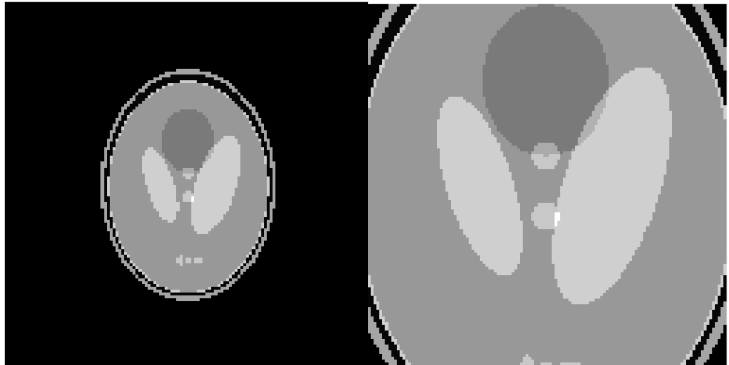

Two simulated images of the same brain-like object but different FOVs are shown above. The image with the higher FOV (left) covers the entire object imaged. The image with the small FOV (right) is focused on one part of the image while the rest is discarded.

e. **The relationship between FOV, Voxel Size, and Matrix Size**: Since the following three variables are dependent on each other, they cannot all be freely set. This is similar to a simple y=mx+z function. You can set two of the three variables freely and solve for the other but you cannot set the value for all three variables and have the equation be true everytime. Similar to that, we can change two of either FOV, Voxel size, or Matrix Size and solve for the other variable. The equation relating all three variables is:

*Field-of-View = Matrix Size x Voxel Size.*

f. **Windowing**: Windowing affects the image contrast by filtering out certain signal intensities. When you window an MRI image, you are selecting which signal intensities you want to view and which you want to ignore. The window is defined by two values: the "min intensity" and the "max intensity". When an image is windowed, all values <= min level are set to zero while all values >= max level are set to 1. The values in between min level and max level are rescaled so they range from 0 to 1. The width of the window is defined as:

*window width = max level - min level*

While the "level" of the window is defined as their midpoint:

*level = 0.5 x (max level + min level)*

The wider your window, the more intensity values are present and the more contrast you can cover, but smaller contrasts are less apparent. The opposite is true for when the window gets narrower as fewer intensity values will be selected and tiny intensity differences will appear more dramatic.

g. **Voxel size vs. true resolution**
Voxel size is often representative of resolution, but not always. One definition of true spatial resolution is how close together (say, in millimeters) two thin lines can get before they can no longer be told apart. In fact, with a technique called zero-filling, we can generate images of very large matrix sizes from images of smaller matrix sizes. In this process, the voxel size is reduced, but true resolution is not improved because the new image contains no more information than the old, pixelated image. One example is shown below:

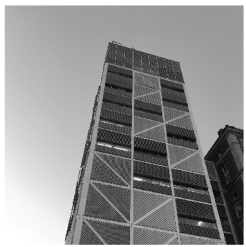 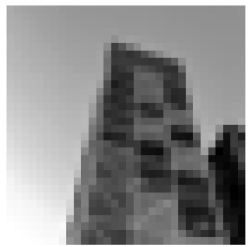 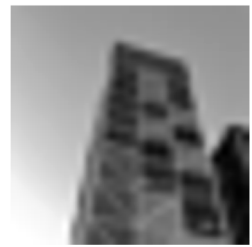

1024 x 1024    Resized 32 x 32    Zero-filled back to 1024

When the 32 x 32 image is zero-filled back to 1024 x 1024, the details are still lost. The information we missed out on when acquiring fewer pixels from the beginning are NOT recovered by the zero-filling. It merely "smoothes out the pixels" but does not improve true resolution at all.

4. Lab Procedure

Image the star tube with 3 Field-of-View settings: largest FOV(pick the largest value of those available), smallest FOV(pick the smallest value of those available), and perfect FOV (re-run the experiment until you find the perfect amount). Fill out the below table.

|  | Biggest FOV | Smallest FOV | Perfect FOV |
|---|---|---|---|
| Value |  |  |  |
| Impact on Resolution |  |  |  |

  i. Sketch and describe what you see for biggest, smallest, and perfect FOV

    1. Big:

    2. Perfect:

    3. Small:

ii. Keeping the perfect FOV, change the matrix size and describe the images. Which matrix size produced the highest resolution?

|  | Matrix size = 16 | Matrix size = 32 | Matrix size = 128 |
|---|---|---|---|
| Resolution |  |  |  |
| Image features |  |  |  |

Image the star tube a few different matrix sizes and zero-filled sizes; fill out the below table with your findings.

| Voxel size (mm) | Matrix Size | Zero-filled to | Observations |
|---|---|---|---|
|  |  |  |  |
|  |  |  |  |
|  |  |  |  |
|  |  |  |  |
|  |  |  |  |

iii. Describe what you see as you change each of the following:

1. Matrix size:

2. Voxel Size:

3. Zero-fill size:

iv. How would actual resolution be affected with the following:

1. Increasing matrix Size while keeping FOV the same:

2. Decreasing voxel size while keeping matrix size the same:

3. Keeping FOV and matrix size and increasing zero-filled size:

Adjust the windows on the image display to explore its effects. Fill out the table with your findings about intensity values. The bullet points below suggest some specific combinations to try:
- Min = 0, max = 1
- Min = max = 0.5
- A high minimum (0.31-0.5) with low maximum (0.51-0.8)
- A low minimum (0.1 - 0.3) with a low maximum (0.51-0.8)
- A high minimum (0.31-0.5) with a high maximum (0.81-1)
- A narrow window
- A wide window

| Experiment | Min intensity (0-1) | Max intensity (0-1) | What can you see? | What is missing? |
|---|---|---|---|---|
| 1 | | | | |
| 2 | | | | |
| 3 | | | | |
| 4 | | | | |
| 5 | | | | |
| 6 | | | | |
| 7 | | | | |

Take-Aways:
1. What happened when the min value was decreased while the max value stayed the same?
    a. Ans.
2. What happened when the max value was increased while the min value stayed the same?
    a. Ans.
3. What happened when the width of the window was increased?
    a. Ans.

5. Questions

a. What combination of matrix size and FOV leads to the highest resolution?
    i. Matrix size = 16, FOV = .16 mm
    ii. Matrix Size = 18, FOV = .36 cm
    iii. Matrix Size = 128, FOV = .25 cm
b. Why does acquiring smaller pixels lead to better resolution?
    i. More pixels means a larger range of gray values are possible
    ii. A small pixel contains more information than a large pixel
    iii. Pixel amount and size have no effect on resolution
    iv. Each pixel represents a smaller amount of space
c. What min/max window settings would allow the most visual contrast to show up between the two circles (numbers indicate gray level)?

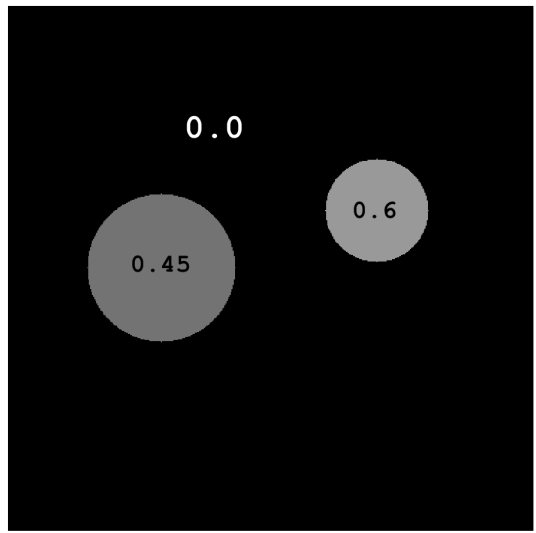

    i. Min: 0.01 Max: 0.99
    ii. Min: 0.2 Max 0.4
    iii. Min: 0.5 Max: 0.5
    iv. Min: 0.3 Max: 0.7
d. Look at the following images and answer the questions:
    i. Which image has the highest Resolution?

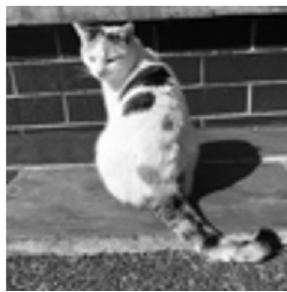 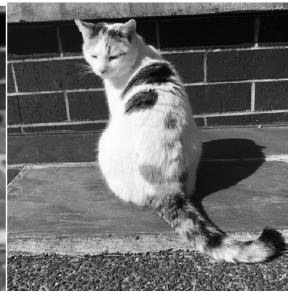 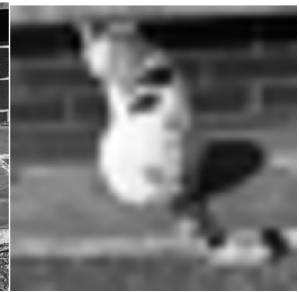

(a)  (b)  (c)

ii. Which image has the tiniest FOV?

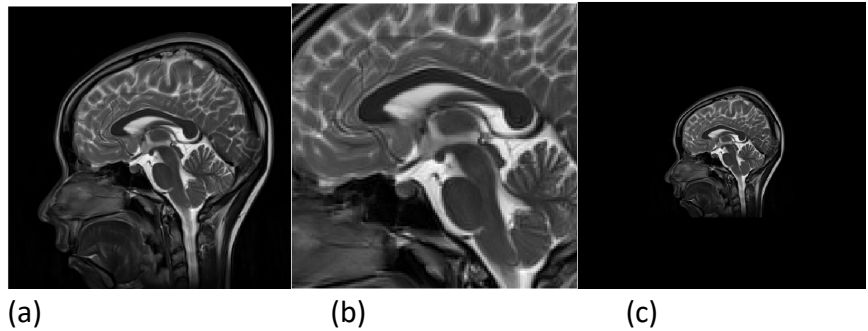

(a)　　　　　　　　　(b)　　　　　　　　　(c)

iii. Which image has the smallest matrix size?

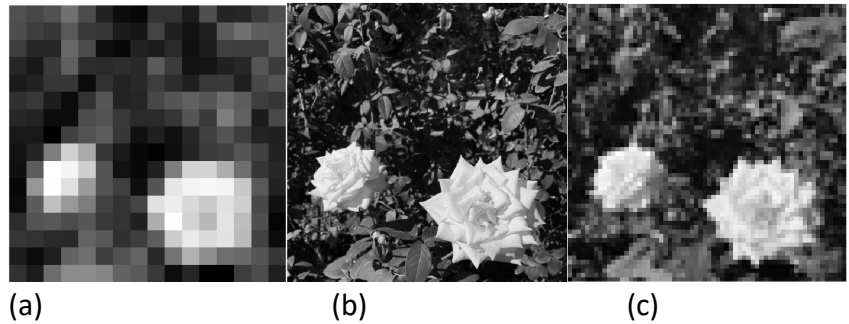

(a)　　　　　　　　　(b)　　　　　　　　　(c)

e. What is the difference between a pixel and a voxel?
   i. Voxels are larger than pixels and hence contain more information
   ii. Voxels are sized by volume while pixels are sized by area
   iii. Voxels make up color images while pixels make up monochrome images
   iv. Only MR images are made of voxels; all other medical images are made of pixels

f. What does zero-filling this image of a pineapple do?

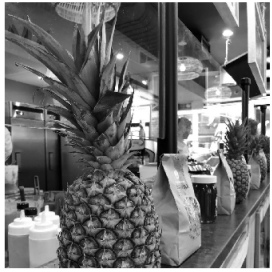 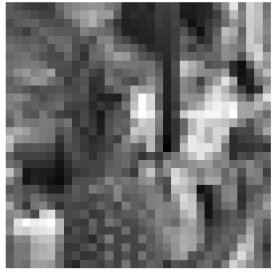 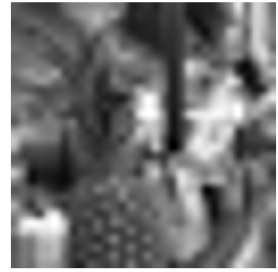

   (A) Original        (B) Reduced matrix size    (C) B, Zero Filled to original

i. It makes each pixel represent a smaller section of space
ii. It makes the fruit more delicious
iii. It sets the background values to zero and makes the pineapple stand out
iv. It improves the true resolution of the image before zero-filling

# Game 2 Lab Manual: K-space magiK

**Why?**
MRI gets its data from the so-called "k-space", which is a special representation of the image we acquire. Just like sheet music breaks down the musical signal (the audio waveform) to its frequency components (notes), k-space shows us what spatial frequency "notes" or harmonics exist in the image. Because there is a one-to-one relationship between the image and k-space, we can travel freely between the two without losing information. Changing and deleting parts of k-space will have interesting effects on the image, which are very relevant for MRI because we don't get a perfect k-space all the time as random noise and other factors start corrupting it. In this game, we will tweak our data in many ways and explore the magical relationship between the two domains!

**Materials**
- Water phantom
- Star phantom
- Letter phantoms
- Brain sample tube

**Background**
1. Key terms
    - Signal
    - Sampling
    - Spectrum
    - Fourier transform
    - Image space
    - K-space

2. Basics
    An image can be seen in more than one way. Just like a piano sonata can be represented either as a WAV file (a time domain signal) or discrete notes on a sheet music, an image can be in the form for human eyes as a grayscale map or as its spatial frequency domain, or k-space. Any 2D image (MR image, photo, drawing) can be converted to a k-space, and any 1D signal can be converted to a spectrum. The conversion process is essentially the same.

3. Explanations
    Signal: a signal is a function representing some processes we are eager to investigate. It can be discrete or continuous, categorical or interval, and in 1D or

2D or 3D or more dimensions. Examples: ECGs, brain waves, average June temperature of the past 10 years in your city... In this game, "signal' is used specifically for a 1-dimensional continuous wave/curve in time. For example:

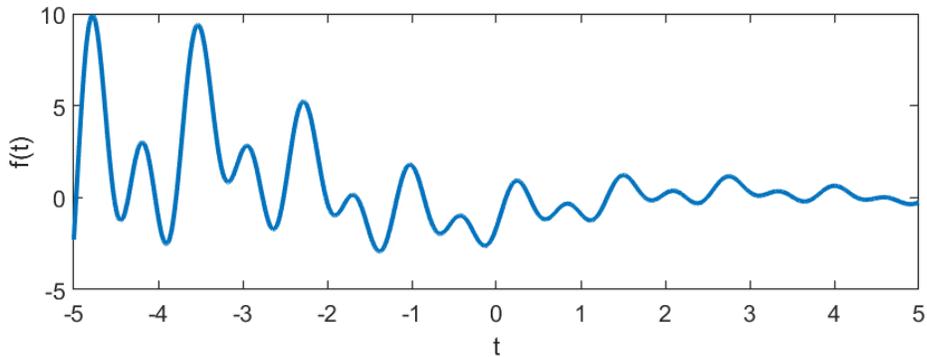

Sampling: This refers to the process of gathering data at selected locations in k-space or in time. In discrete sampling, we get an approximate idea of a continuous signal by looking at it at a limited number of times. Below, the signal above is sampled at two different rates and the resulting discrete signals look quite different. We want to sample just enough to get a fair representation of our signal.

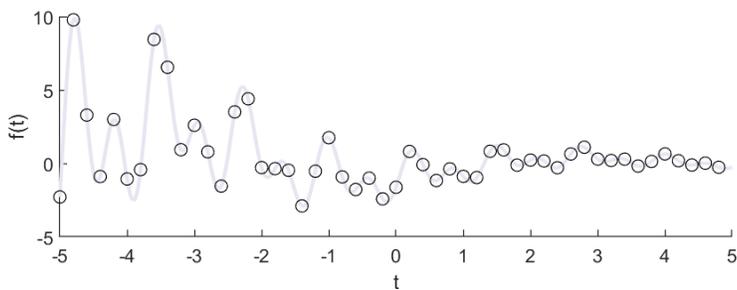

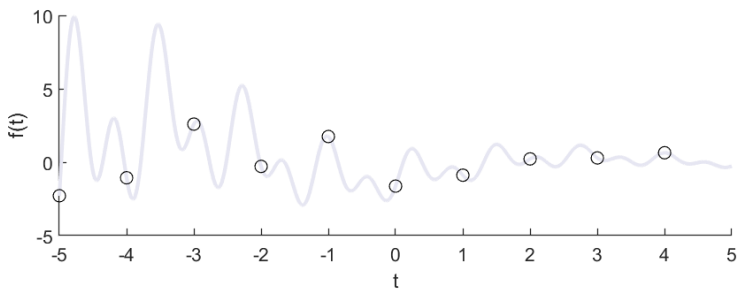

Spectrum: this is a curve that shows you how much of each frequency is in a signal. For example, the following signal is a sum of 3 different sinusoidal waves:

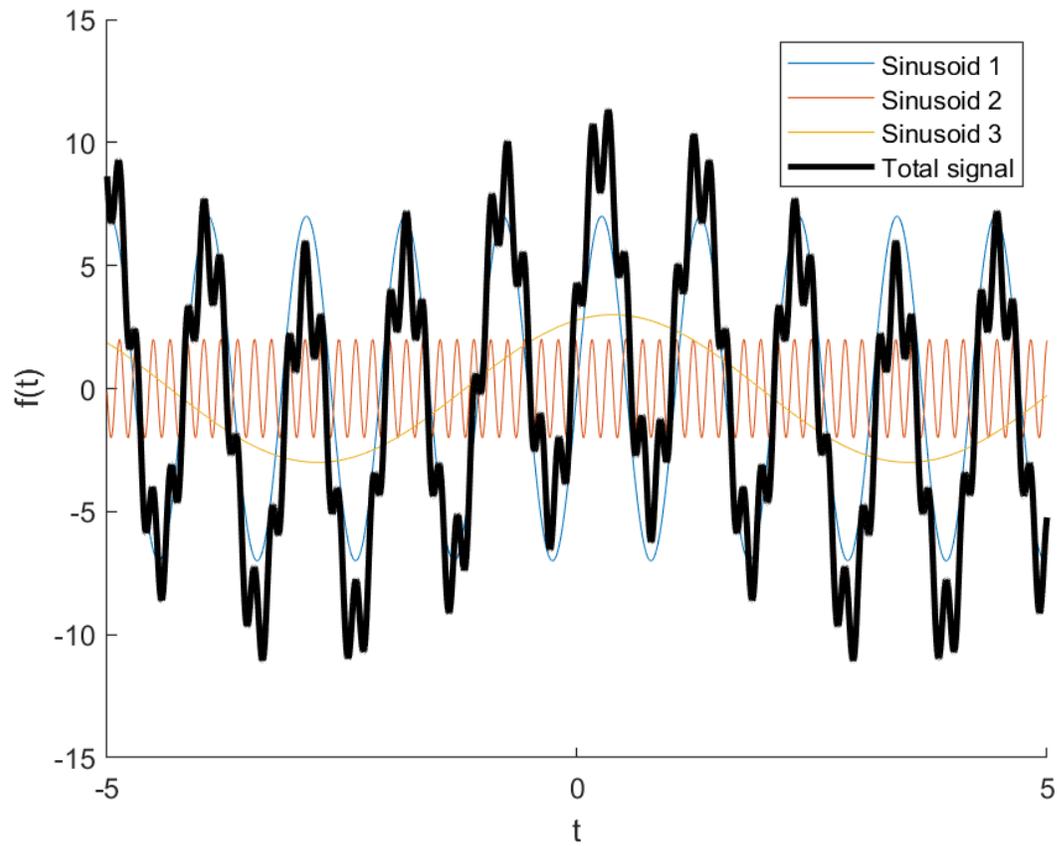

Its spectrum is shown below. As you can see, the three waves are represented as three pairs of peaks on the spectrum, each corresponding to a distinct frequency as marked by the horizontal axis.

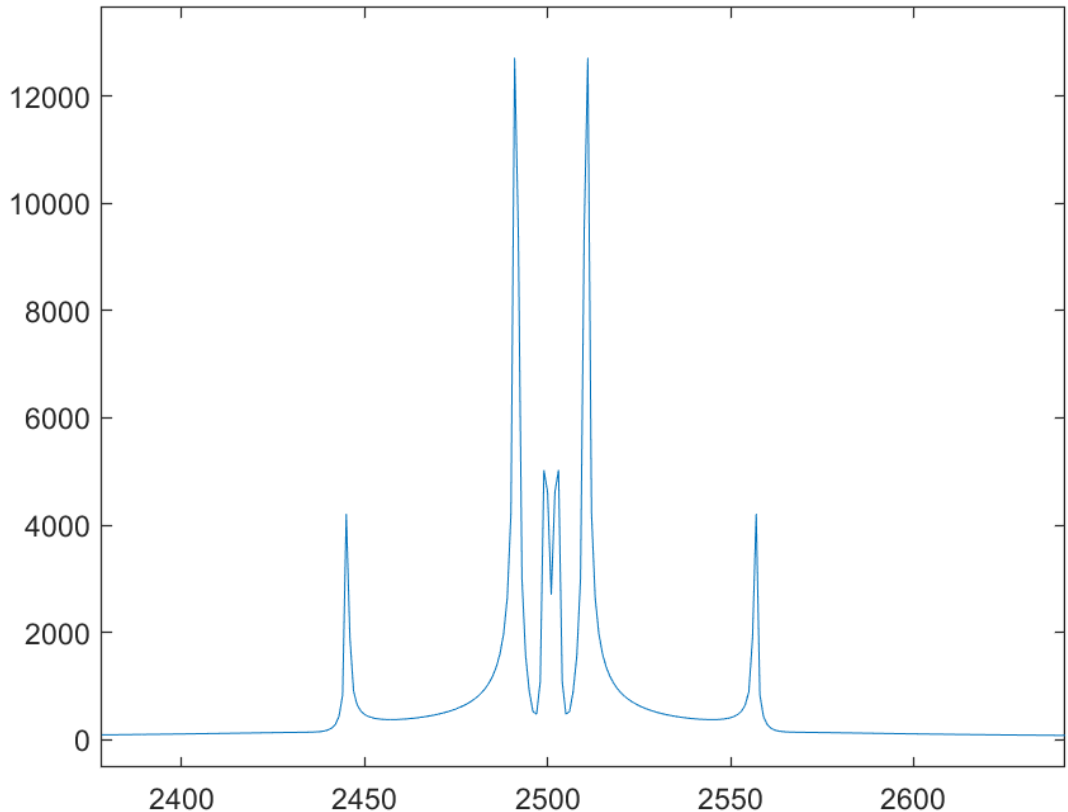

Each pair of peaks on the spectrum corresponds to a sinusoidal wave on the signal. Higher peaks means the component has larger amplitude, and peaks closer to the center means the component has lower frequency.

Fourier transform (FT) is a mathematical operation that converts a **signal** to its **spectrum**. This process is one-to-one: we can apply the inverse Fourier transform to the spectrum and get back the same signal. The amount of information is preserved before and after the transformation. The equation for a FT and an inverse FT (IFT) is shown below. It really is just an integral with complex exponentials thrown in!

Forward FT:
$$X(f) = \int_{-\infty}^{+\infty} x(t) e^{-i2\pi f t} dt$$

Backward IFT:
$$x(t) = \int_{-\infty}^{+\infty} X(f) e^{+i2\pi t f} df$$

Some interesting properties of the FT are summarized below:
1. The FT of the sum of two (or more) signals is equal to the sum of the FTs of those two (or more) signals.
2. The FT of the product of two signals is equal to their FTs convolved together (https://en.wikipedia.org/wiki/Convolution), and vice versa.
3. If you multiply a signal by a constant C, the FT of it is also multiplied by C.
4. If a signal gets wider, its FT gets narrower, and vice versa.
5. If you move the signal horizontally (to the left or right), its FT gets a linearly varying complex phase. If you move the FT, the signal gets the same.

You will get to explore some of these properties on the game interface!

Image space is the space explored in Game 1: grayscale maps represented as matrices. Each entry of the matrix represents a point in space and its magnitude tells you how bright the image is at that spot.

K-space is what you get if you perform a Fourier transform on the image space. Because the image space is 2D, we have to integrate in two directions.

$$G(k_x, k_y) = \int_{-\infty}^{+\infty} \int_{-\infty}^{+\infty} g(x,y) e^{-i2\pi(k_x x + k_y y)} dx dy$$

The same properties apply to the 2D FT as described above for the 1D FT.

Spatial frequency: if a sine wave is represented by a pair of peaks on the spectrum, what is represented by a pair of peaks in k-space? This is where spatial frequency comes in. See the first figure below:

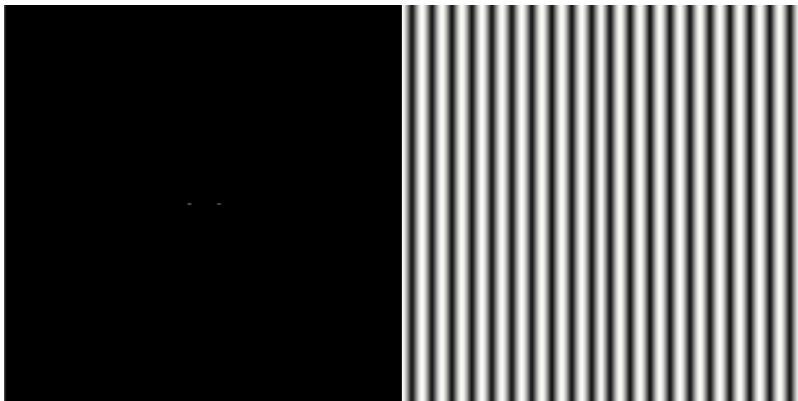

K-space                          image

In k-space, there are two peaks symmetric by the origin (look very closely, or zoom in if needed). The line connecting the peaks is horizontal. In the image, we see stripes. The brightness is not either black or white, but varies along the same horizontal line in the form of a sinusoidal function. Compare this to the next figure:

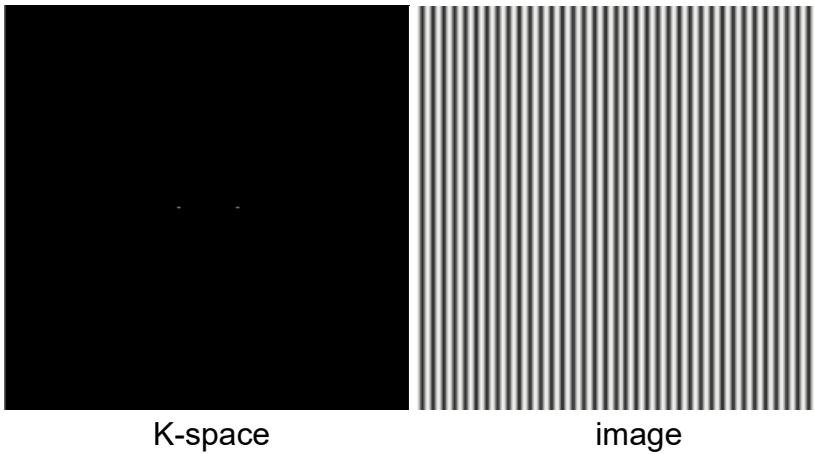

K-space                         image

Here, the peaks are further apart but still oriented along the horizontal line. The stripes are narrower indicating a **higher spatial frequency**. Hence, each spatial frequency corresponds to a 2D sinusoid wave of a certain width. The higher the frequency, the closer the peak gets to the edge of k-space, and the narrower the stripes get because they vary faster in a given amount of distance. Now, observe the last figure below:

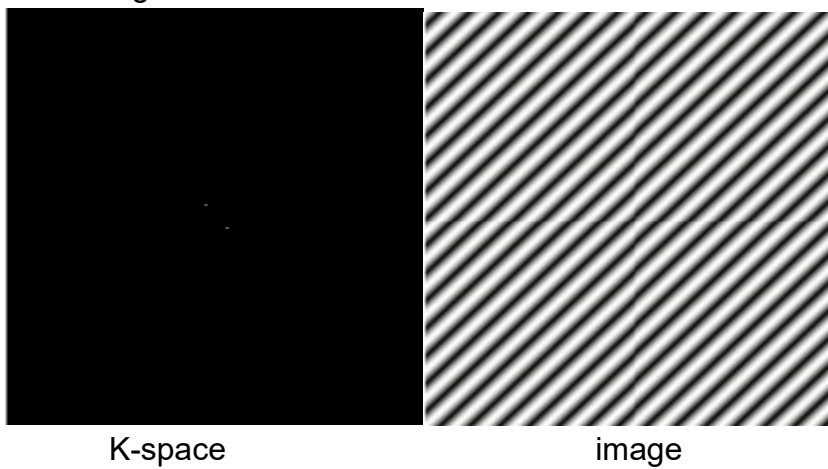

K-space                         image

The peaks, while still symmetric about the origin, have been rotated by 45 degrees. The 2D wave was also rotated and the direction perpendicular to the stripes still corresponds to the line connecting the peaks. Therefore, the 2D spatial waves have both a frequency and a direction defined by an angle between 0 and 180 degrees.

Importantly, the points in k-space and image-space do NOT map one-to-one. Changing one point in k-space will change ALL POINTS in the image space, and vice versa. This is because each point in k-space corresponds to a spatial wave that occupies the entire image. Because of this, many image artifacts in MR are not straightforward, but require some amount of "k-space intuition" to understand. You will further develop this intuition with the Game's GUI.

k-space and MRI acquisition: while the camera and the human eye both speak the language of image space (as projected onto the image sensor in the camera and the retina in the human eye), MRI speaks in k-space - it is the natural place where we get our data. This is because of the following:
1. The protons in hydrogen atoms are tiny magnets that, when placed in a magnetic field, rotates around that field's direction.
2. The stronger this field is, the faster they rotate.
3. A rotating thing can be represented as a complex exponential. For example, the position of a dot traveling on a unit circle counterclockwise at constant speed can be represented as:
$$x(t) = cos(\omega t), \quad y(t) = sin(\omega t)$$
Or:
$$x(t) = Re\{r(t)\}, \quad y(t) = Im\{r(t)\}$$
Where
$$r(t) = e^{i\omega t} = cos(\omega t) + i\, sin(\omega t)$$
Is a complex function representing the position function.
4. When we apply an imaging gradient field (one of the three key magnetic fields that the MRI scanner uses), the complex exponentials get their frequencies modulated in such a way that the signal **is** the Fourier transform of the image. The signal in general is a spatial integral of all the tiny rotating magnets, so this process **physically** performs a Fourier transform (with the protons themselves being the complex exponential terms) !
5. If we apply the gradients correctly, each point in time gets mapped to a different point in k-space, and we know this mapping.
6. By sampling the signal at different times, we can fill up k-space.
7. Lastly, we perform an inverse 2D FT to convert the k-space back to the image. This step is called image reconstruction.

**Lab procedures**
1. Explore 1D transforms

a. Select the "sine wave" signal type on image/signal presets and press "Get signal". Describe the signal in your words:

b. Press "Forward" to perform a Fourier transform. The right chart now displays a spectrum of the signal. Describe the spectrum.

c. Try changing each of the parameters, generate the signal, and get its spectrum each time. Describe what each parameter does to the signal and to the spectrum:

|  | Signal | Spectrum |
|---|---|---|
| Vertical Scale |  |  |
| Horizontal Scale |  |  |
| Shift |  |  |
| Phase modulation |  |  |

d. Experiment with other options in the "select signal" dropdown.
e. Experiment with options in the "select spectrum" drop down on the right. Use the "backward" button to perform an inverse Fourier transform, which recovers the signal from its spectrum. What happens when you press Forward and then Backward or vice versa?

f. In the middle, go to the "Draw" tab and draw any curve you like. Press "Use" and then "Get signal" or "Get spectrum" to load it onto the panel you want. Perform transformation of your signal and spectrum. Use the drawing board and the preset signals to explore at least one of Properties 1-5 of the Fourier transform. Record your findings below with sketches of the signals and spectra you used.

| Property # | Signal | Spectrum | Findings |
|---|---|---|---|
|   |   |   |   |
|   |   |   |   |
|   |   |   |   |

2. Explore 2D transforms
    a. Now we can look at the 2D analog of the same process. Go to the "2D Image" tab and choose any image you like. Press "Get Image" and then "Forward" to look at the k-space. Repeat it for a few images.

       What do the k-spaces have in common?

    b. Get the image of a 2D sine wave and generate its k-space. What do you see? You might have to zoom in to see the details of this one. How does this compare to the 1D sine wave?

c. Try changing each of the parameters below, generate a new image, and get its k-space each time. Describe what each parameter does to the image and to the k-space:

|                  | Image | K-space |
|------------------|-------|---------|
| Rotation         |       |         |
| Horizontal Scale |       |         |
| Shift            |       |         |
| Phase modulation |       |         |

d. Perform the same steps in c, but using the Backward transform on the "Double spike" 2D k-space preset (press "Get K-space" to generate the k-space first).

|                  | K-space | Image |
|------------------|---------|-------|
| Rotation         |         |       |
| Spike separation |         |       |

e. Explore other images and k-space options and perform multiple backward and forward transforms.

f. In the middle, go to the "Draw" tab and draw any image you like. Press "Use" and then "Get image" or "Get k-space" to load it onto the panel you want. Perform transformation of your signal and spectrum. Use the drawing board and the preset signals to explore at least one of Properties 1-5 of the Fourier transform in 2D. Record your findings below, with sketches of the images and k-spaces for each experiment.

| Property # | Image | K-space | Findings |
|------------|-------|---------|----------|
|            |       |         |          |
|   |   |   |   |
|   |   |   |   |
|   |   |   |   |

3. Perform k-space magik!
    a. Load one of the MRI images and generate its k-space.
    b. Go to the "Sampling" tab in the middle. The square represents k-space and the four lines slice it up. The lighter part will be preserved and the darker part will be erased from our k-space. Press "APPLY" to see the slicer applied.
    c. Move the four lines closer to the middle of k-space so only a small portion of light gray remains, and press apply. Transform the restricted k-space back to image space. How did the image change? Why? Can you relate this to the sine wave experiments above? State your conclusion:

    d. Re-load the image and generate a fresh, complete k-space; then, press "Invert" in the slicer and "Apply" to block out the center but preserve the periphery of k-space. Transform backwards. What do you see now? Describe your findings:

e. Explore the slicer to section out different parts of k-space and see its effects. Moreover, you can also use the "Erase" tab to block out more interesting shapes and see what it does to k-space! Describe your findings in the table:

| K-space blockout (describe or sketch) | Effects on the image |
|---|---|
|  |  |
|  |  |
|  |  |
|  |  |

4. Free exploration suggestions (optional)

Note: On the "upload" panel, you can upload an image and generate its k-space. Use "Erase" or "Slicer" to apply different filter effects onto your image and then transform backwards. "Recover" reloads the complete image. You can use the toolbar on top to save a local copy of the resulting image.

   a. Explore the effects of undersampling factors on the "sampling" panel.

   b. Using the properties of k-space, plan out and create your own art by drawing an image, converting it to k-space, manipulating it with "erase" and "sampling", and exporting it.

   c. Using a photo you have taken and knowledge of k-space, style it in three different ways by changing its k-space and export them.

   d. Compose a haiku or a short free-form poem about k-space using what you learned today.

**Questions**
Q1. Which statement below is incorrect about the Fourier Transform?
   A. It is an irreversible process because you cannot figure out the original signal from the transformed signal.
   B. Mathematically, it can be performed as an integral with complex exponentials over time or space.
   C. The FT of the product of two signals is not necessarily the product of its FTs.
   D. The output of the inverse Fourier transform can be entirely real (i.e. without imaginary components)

Q2. What does k stand for in the term "k-space"?
   A. Key imaging variable
   B. Kinetic energy conversion
   C. Spatial frequency
   D. Coordinates of reconstruction

Q3. What statement about k-space is correct?
   A. The middle of k-space represents the edge information and small details on the image
   B. The periphery of k-space represents the overall signal level of the image

C. K-space is artificially created from the acquired image so we can manipulate the image contrast
D. K-space is directly sampled and the image is created afterwards in an extra step

Q3. What happens if I slice the k-space in the following way:

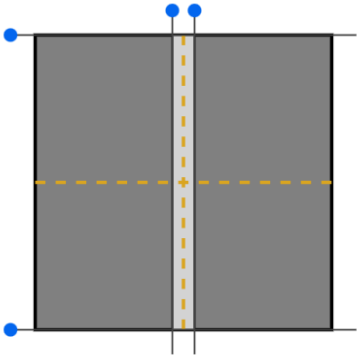

A. Vertical edges will get blurry
B. Horizontal edges will get blurry
C. Only vertical edges will be visible
D. Only horizontal edges will be visible

Q4. What is incorrect about spatial frequency?
A. Low spatial frequencies corresponds to information in the center of image space
B. Each point in k-space tells you the amplitude of one spatial frequency
C. High spatial frequencies corresponds to thinner stripes
D. In 2D, spatial frequency has a direction along which the wave amplitude changes

# Game 3: Brains, please!
## Why?

Now that we have a firm understanding about how images work, it is time to expand our vision with some "filters" we apply to the scan to look for certain properties. Like instagram and snapchat filters that exaggerate certain aspects of the face, MR "filters", or contrasts, exaggerate components in the scan like fat or water. Learning about how to control these contrasts can help us see the brain in many lights and identify things that we can't otherwise.

## Materials

- Brain specimen tube

## Background:

1. <u>Key Terms</u>
    a. Brain tissue types
    b. RF Pulse
    c. FA
    d. TE
    e. TR
    f. T1, T2, PD
    g. T1w, T2w, PDw

2. Basics

    Like any other part of the body, the brain has many types of tissues with different physical properties. When using the MR scanner, we often rely on them to tell apart brain regions and look for areas where these properties change because of diseases. To highlight specific tissues, we can use a T1-weighted, T2-weighted, or PD weighted scan.

3. Explanations
    - **Brain tissue types:** There are 3 major types of brain tissues: **Cerebrospinal fluid (CSF)**, **White Matter (WM)**, and **Gray Matter(GM)**. Cerebrospinal fluid is the liquid content of brain ventricles. It flows in and around the brain to absorb impact from injuries to the skull and provide nutrients. White matter effects learning, distribution of action potentials, and communication between different regions of the brain. It's called white matter because its neuronal axons are covered in a protective fatty sheath which gives it a white color. Gray matter mainly receives incoming information and regulates outgoing information.

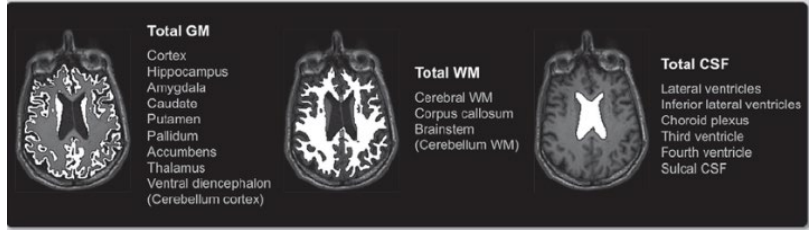

https://www.researchgate.net/profile/Daniel-Ferreira-4/publication/283773154/figure/fig1/AS:309012361957376@1450685696573/BV-CSF-index-Total-grey-matter-GM-volume-total-white-matter-WM-volume-and-total.png

- **RF Pulse**: this is a short-lived magnetic field that tips the protons off their main magnetic field axis so that they are at an angle to it. How much they get tipped depends on the strength and duration of the RF pulse.

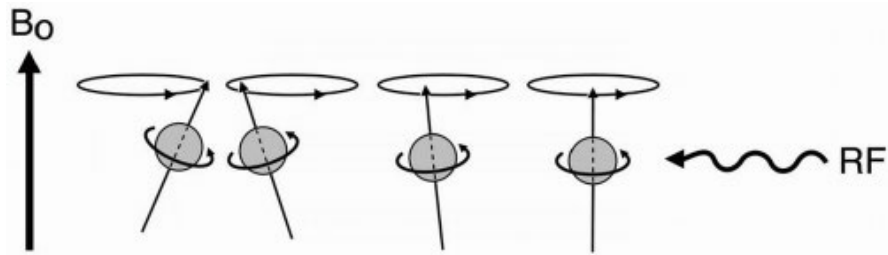

https://radiologykey.com/wp-content/uploads/2016/05/C3-FF1-2.gif

- **Flip Angle (FA)** is the net rotation angle of the magnetization when an RF pulse is applied. A FA value of 0 signifies no change, while a 180 degree FA value signifies a rotation of 180 degrees so the magnetization is pointing straight down. At low flip angles, the MR signal is roughly proportional to the flip angle. This no longer holds at higher flip angles, and the signal dependence on flip angle gets more complicated and depends on the next two MR parameters.

-   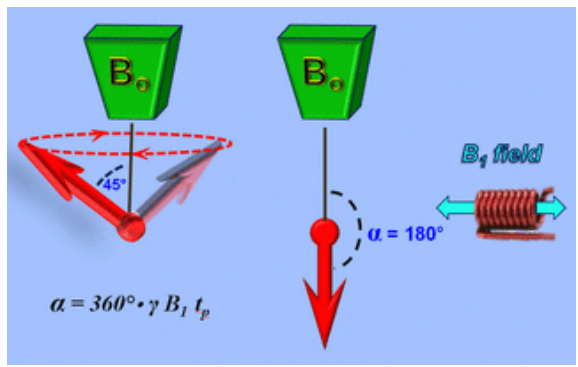

- **Echo Time (TE)** is the time difference between the time when the RF pulse arrives at the target to the time when the signal bounces back and hits a peak, forming what's called an echo.

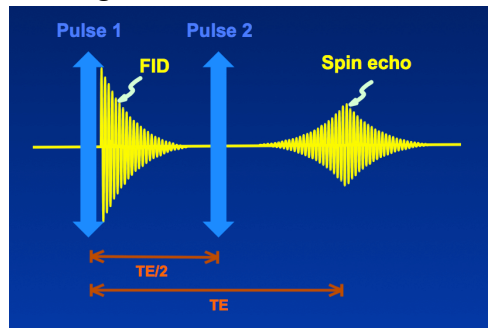

http://mriquestions.com/uploads/3/4/5/7/34572113/_7707793_orig.gif

- **Repetition Time (TR)** is the time difference between one set of RF pulses (a 90 degree pulse and an 180 degree pulse in the example below) and the next. It is called "repetition" because we excite the spins in exactly the same way between the two intervals. We need to repeat the excitation because we only have time to get partial information within each TR. Combining signals from multiple TRs gives us a complete image.

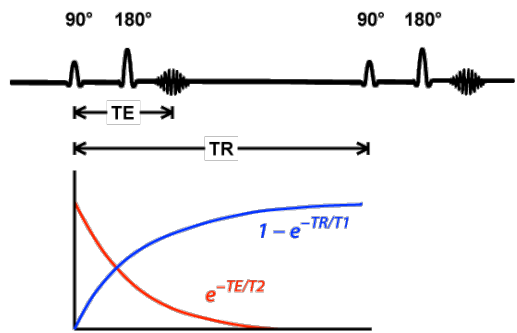

https://mriquestions.com/uploads/3/4/5/7/34572113/8806734_orig.gif

- **Longitudinal Relaxation Time (T1)**, **Transverse Relaxation Time (T2)**, and **Proton Density (PD)** are the main physical properties measured by MRI.
    - T1 and T2 have units of seconds or milliseconds and decide how fast signals "relax" back into their equilibrium value.
    - PD refers to the relative density of protons, so it measures how much signal we can get from the same volume of tissue. All other things being equal, the higher a tissue's PD is, the brighter it looks on the image.

The table below shows typical T1, T2, and PD values for the three brain tissue types at a main field strength of 1.5 Tesla.

|  | GM | WM | CSF |
|---|---|---|---|
| **T1 (ms)** | 1130 | 750 | 1940 |
| **T2 (ms)** | 119 | 87 | 230 |
| **PD (relative)** | 1.04 | 0.95 | 1.02 |

https://campar.in.tum.de/pub/buonincontri2018ismrmrep/buonincontri2018ismrmrep.pdf

Table 1: Typical brain tissue parameter values at 1.5T

- **T1-, T2-, and PD-weighted (T1w, T2w, PDw)**: You can think of these as 3 "filters" for MR images. In general, T1w highlights tissues with short T1, T2w highlights tissues with long T2, and PDw highlights tissues with high PD.

  - T1w is known for highlighting fat, proteins, melanin, and areas of breakdown in the blood-brain barrier, indicating inflammation.
  - T2w highlights liquids such as CSF in the brain and detects deoxygenated hemoglobin, methemoglobin(a type of hemoglobin that can't give out oxygen to tissues), or hemosiderin (caused by blood leaking out of capillaries) in lesions and tissues.
  - PDw highlights tissues with high concentration of protons and can be used to assess joints: it offers distinct contrast between fluid, cartilage, and bone

| Contrast | TE | TR |
|---|---|---|
| T1w | Short | Medium |
| T2w | Medium | Long |
| PDw | Short | Long |

Table 2: How to set TE and TR to get different contrasts

## Lab procedures

1. Insert the brain specimen tube and perform scanner calibration.
2. Start off with Panel 1. There are three imaging options present: T1-weighted, T2-weighted, and PD-weighted. Select each of the options and click run to acquire an image . How are the scans different from each other? Note down below on relative signal strengths and what each of the scans highlights.

| Weighting Calibration | Tissue with highest Signal | Tissue with medium Signal | Tissue with lowest Signal | What does the scan highlight? |
|---|---|---|---|---|
| T1 Weighted | | | | |
| T2 Weighted | | | | |
| PD Weighted | | | | |

Answer the question at the bottom of the panel to continue.

3. After you correctly answer the question on Panel 1, you will be prompted to continue to Panel 2. Here you will adjust TR, TE, and FA values to observe how the scan changes. Also observe how certain TR and TE values produce the same results as T1 and T2 weighted scans. Pick 5 sets of TR, TE, and FA values to fill out the table below.

| | TR (ms) | TE (ms) | FA (deg) | Observations |
|---|---|---|---|---|
| Set 1 | | | | |
| Set 2 | | | | |
| Set 3 | | | | |
| Set 4 | | | | |
| Set 5 | | | | |

Describe what you see as you change each of the following parameter while fixing the other two:

- TR value:
    - As you increase this value, is the contrast increasing or decreasing?
        - Ans:
    - How does contrast change when the TR value is large and the TE value is short?
        - Ans:
- TE value:
    - As you increase this value, is the contrast increasing or decreasing?

- How does contrast change when the TR value is short and the TE value is large?

- FA value:
  - What do you notice when the FA value gets larger?

  - How does the signal dependence on FA change as you increase TR?

Answer the question at the bottom of the panel to continue.

4. After you correctly answer the question on Panel 2, you will be prompted to continue to Panel 3, revealing a graph with signal strengths for CSF, WM, and GM. How can you use this graph to produce different contrasts? Fill out the table below based on what you learned earlier in the lesson.

|  | TR (ms) | TE (ms) | FA (deg) | Goal |
|---|---|---|---|---|
| Set 1 |  |  |  | High signal in CSF and low in GM |
| Set 2 |  |  |  | Very little contrast between GM and WM |
| Set 3 |  |  |  | GM > WM > CSF |

## Questions

1. Which set of parameters gives you the most T1 weighted scan?
    a. TR = 4 s, TE = 2 ms
    b. TR = 5 s, TE = 500 ms
    c. TR = 900 ms, TE = 5 ms

d.  TR =  800 ms, TE = 500 ms

2. What does this PD weighted scan highlight?

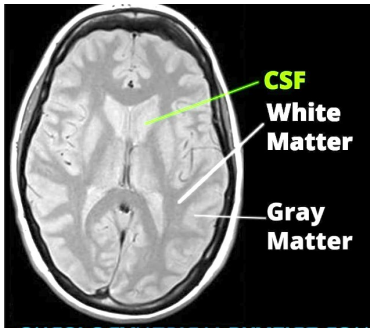

https://oncologymedicalphysics.com/wp-content/uploads/2020/01/Proton-Density-Weighted.jpg

   a. It highlights tissues with the highest density of protons
   b. It highlights tissues with the lowest density of protons
   c. It highlights tissues only at a specific density that the imager can specify
   d. It highlights tissues with no protons at all

3. How would substances with small T2 values appear on a T2-weighted scan?
   a. It would appear bright
   b. It would appear dark
   c. It would appear larger than its physical size
   d. It would appear smaller than its physical size

4. What is a T1 weighted image useful for?
   a. Assess the relative densities of brain tissues
   b. Detecting breakdown in the blood-brain barrier
   c. Detecting deoxygenated hemoglobin
   d. Visualizing the ankle joint

5. Based on Table 1, predict what tissue type (GM/WM) will be highlighted with each filter or MR contrast. Circle the correct tissue type for each contrast:
   a. T1w:  GM   WM
   b. T2w:  GM   WM
   c. PDw: GM   WM

# Game 4 Lab Manual: Fresh blood

**Why?**
MRI is versatile because it can look at not only shapes and anatomies but also physiological processes. These include processes from localized brain activation (the BOLD signal for functional MRI) to changes in chemical composition in specific brain areas (Magnetic Resonance Spectroscopic Imaging). In this game, we look at a simple process happening all the time in our bodies - blood flow. MRI also has specific tools to see flow and map out the shapes of blood vessels. This opens the possibilities of diagnosing injuries, plaques, aneurysms, and obstructions along the vessels. We will explore two modes of flow imaging: "bright blood" and "dark blood".

**Materials**
- Flow phantom
- Plastic tube
- Syringe
- Water reservoir
- Water refill

**Background**
1. Key terms
    - Magnetic Resonance Angiography
    - Spoiled Gradient Recalled Echo
    - Steady state
    - Spin Echo
    - T2/T2* decay
    - Maximum/minimum intensity projection

2. Basics
   Water is a main component of blood and the proton nuclei in water ("H" in $H_2O$) generate MRI signals. These signals cannot be detected by default and must be produced by RF pulses which excite signals only within a certain slice (like a thin slice of bread) inside the body. Most imaging methods use two or more pulses spaced out in time to ensure there is enough signal throughout the scan. When the water protons move out of the slice in between pulses, however, they experience fewer pulses than normal. This either increases or decreases their signal compared to unmoving tissue, depending on the number, timing, and strength of the RF pulses. This difference in signal is all we need to see a visual contrast and trace the blood vessels.

3. Explanations
   <u>Magnetic Resonance Angiography</u> refers to any MR imaging method that generates a contrast between blood vessels and other tissue. This contrast can be generated by distinguishing blood from other materials in multiple ways. For example:
   
   (1) Using blood's inherent properties - for example, T1 - to generate a contrast against the background;
   (2) Using the fact that blood is flowing in or out of the imaging slice;
   (3) Using contrast agents injected into the blood to change its properties and enhance contrast.
   
   The figure below shows examples of MRA images [1]. You can see the twists and turns of the brain blood vessels. You may also notice that they can be either brighter ("bright blood") or darker ("dark blood") against the background.
   
   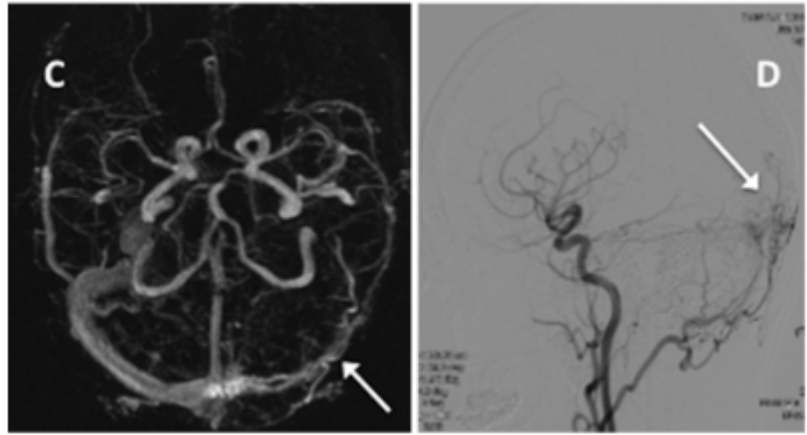
   
   <u>Spoiled Gradient Recalled Echo (SPGR):</u> This is an example of a fast imaging sequence. It uses a train of weak, identical RF pulses with very short spacing in between. The figure below shows all the magnetic field waveforms associated with one pulse [2]. The portion inside the dash lines is repeated at regular intervals (say, once every 10 milliseconds)

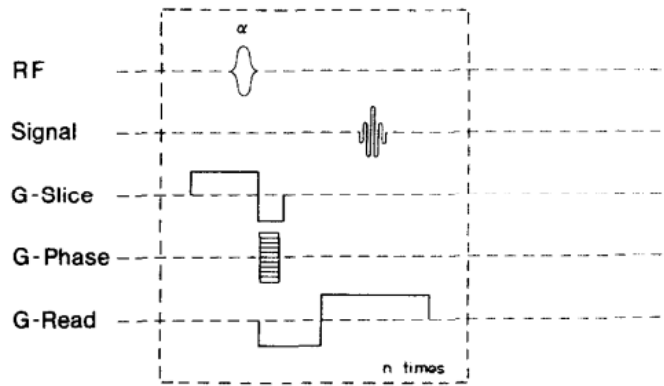

When you represent each RF pulse (shown on the topmost row) as a vertical line, the sequence looks like this:

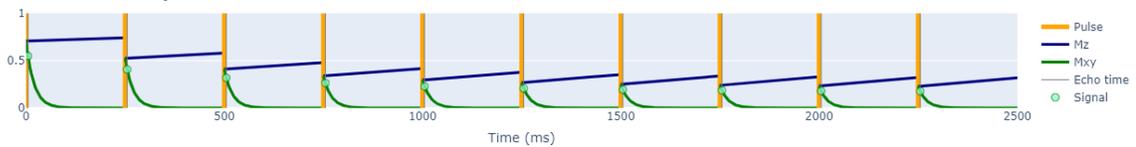

Here, the orange lines represent the many pulses while the blue and green lines represent vector components of the magnetization, over time.

Steady State: After each RF pulse in SPGR, the z component of the signal, Mz (see Game 5), is *reduced by the same fraction*. This component recovers incompletely across the time interval until the next pulse (through the $T_1$ process - see Game 6; also see the blue line in the plot above). At the next pulse, there is a slightly lower initial $M_z$ that again is reduced by the same fraction. Repeat this process a few times, and you will see that the $M_z$ *right before each pulse* stabilizes to a lower value. This is called the steady state value and it ends up being proportional to the signal strength.

Time-of-flight (TOF) effect: RF pulses only excite a limited slice of the body. When protons move *across* that slice, they end up experiencing fewer pulses than unmoving tissues. This is the case for blood, and when the blood flow is fast enough and the slice thin enough, they do not get enough pulses to reach the steady state and emit the higher signals of the first few iterations (non-steady state). This causes blood to look brighter on SPGR images!

Below is a preview of what happens. The horizontal axis indicates position along the slice and blood flows through the central vessel into and out of the slice (orange portion). Because SPGR divides time into regular intervals by having one RF pulse per interval, the slice also ends up being divided into a number of segments, each of which having experienced a different number of RF pulses. The new inflow ("fresh blood") on the left appears brighter because it has only experienced one pulse, while the older blood that has stayed in the slice for longer is darker because it has experienced 10 pulses. The width of each segment is equal to how far blood travels between any one and the next RF pulse. Static tissue does not flow, so all the signal segments would have experienced 10 pulses in the same time and all appear dark. Therefore, blood flow slices generate more signals on average.

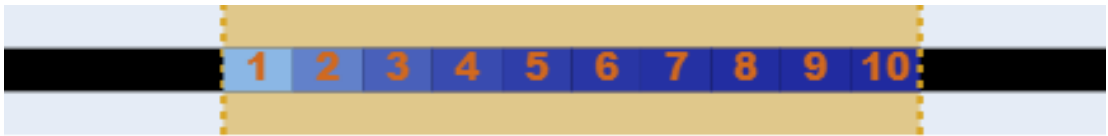

T2/T2* decay

As you will explore more in Game 6 (see lab manual), $T_2$ decay is the reduction of $M_{xy}$ (the component of the magnetization vector in the x-y plane) over time due to spins getting out of phase with each other due to random variations in the local $B_0$ field they experience. This causes an exponential decay of signal from its maximum towards zero. In fact, the $T_2$ decay models an ideal case where the macroscopic field strength is uniform across the sample. In reality, there are spatially varying but time-constant fields in addition to the microscopic random variations. What this means is that the final signal decays even faster than T2. We call this shorter decay time $T_2^*$. $T_2$ and $T_2^*$ are related by:

$$\frac{1}{T_2*} = \frac{1}{T_2} + \frac{1}{T_2'}$$

What all this means is that there are two components to the magnetic field variation that makes spins get out of phase: a magnetic field difference that randomly changes in time and space and is unknown (T2), and a magnetic field difference in space that remains constant in time (T2'). We can assume that two components are independent of each other.

Spin Echo sequences generate a dark blood contrast. This happens because each spin echo *must* be made by applying two RF pulses one after the other. It goes like this:
1. Play a 90-degree pulse (see Game 5) at time t = 0
2. Waiting for a time period of Δt
3. Play a 180-degree pulse
4. Wait another Δt and acquire signal at time 2Δt

The 180 pulse is very useful because it *reverses* the progress of the spins with different speeds, as described above in the $T_2/T_2^*$ definitions: now the last place is the most ahead and the first place is the most behind. After you let them rotate for another Δt, the vectors all line up again! Because of the way vector addition works, they collectively generate a high signal at t = 2Δt. This is called an *echo* and we call the 180 pulse a *refocusing* pulse because it enables an echo by leveling the playing field among the spins. The animation on the Wikipedia page shows this effect (https://en.wikipedia.org/wiki/Spin_echo#/media/File:HahnEcho_GWM.gif)

One key fact is that the $T_2'$ portion (same in time, changing in space) of the signal loss can be recovered but the $T_2$ portion *cannot* as the spin echo requires the $B_0$ to stay the same during the first time interval and the second so the amount of "catch-up" is correct for spins in each location. The $T_2$ portion cannot be *refocused* because $B_0$ fluctuates randomly throughout the sequence.

Dark-blood effect
Now, we can agree that spins that experience both pulses generate the most signal. Spins that only experienced the 90-degree pulse generate less - they will still have some weak signal because the $T_2^*$ effect reduces the signal but doesn't always cancel it all out. Lastly, spins who only experienced the 180-degree pulse generate no signal whatsoever because an 180-degree pulse simply points the equilibrium (0,0,M0) vector to the opposite direction with no xy component - that is, to (0,0,-M0).

What this means for blood flow is that in a given slice, at least some of it will experience only one of the two pulses. Only a part of it (that is, less than 100%) will experience both pulses and generate a spin echo. In contrast, all spins from static tissue can generate an echo. Based on the ranks we assigned on signal levels (both pulses > 90 only > 180 only), we can say that the signal from blood that flows in/out of the imaging slice is *reduced*. This gives you a "dark blood" look on the image.

Maximum/Minimum Intensity Projection
Once the blood vs. background contrast is generated, different processing steps can help visualize the vessels better. For example, maximum intensity projection can be used to highlight bright signal blood vessels. To perform this, you look at the 3D image volume from a certain angle and for each parallel ray of vision that leaves your eye, you choose the maximum signal to display on a flat surface, or your field of view - and that ends up being signals from the blood vessel if you did the image collection right. Similarly, for dark blood image volumes, you could also choose the lowest signal along the ray of vision. This other method is called a minimum intensity projection.

**Lab Procedures**
1. Explore flow and two contrasts
    a. Push the syringe
       Press the "PUSH" button to start moving the syringe at a constant speed. The syringe is connected to a flexible tube that makes a U-turn inside the imaging test tube. Images are acquired perpendicular to the flow. When the plunger is moving, you can press the "STOP" button to stop it. The speed can be changed by changing the "Flow speed" field, pressing "STOP", and then pressing "PUSH" again.
         - What are the directions of flow with regard to the imaging plane?
         - Name two factors that affect the flow speed within the flexible (thin) tube.
    b. Bright blood contrast
       Set flow speed = 10 mm/s. While the plunger is not moving, press "RUN" on the right side to acquire an image (make sure "contrast type" is set to "bright blood"). Then, start pushing the syringe and acquire another image while there is flow.
         - What is the difference between the two images?

    c. Dark blood contrast
       Set flow speed = 10 mm/s, contrast type = dark blood. Push "transfer parameters". Then acquire one dark blood image without flow and one with flow.
         - What is the difference between the two images?

2. Explore the time-of-flight effect
    a. Observe the time-of-flight diagram
       After pressing the "PUSH" button once, observe the graph below the syringe. On the "BRIGHT BLOOD" tab, there are two plots: the top one showing RF pulses and the magnetization over time, and the bottom one showing flow. There is a time axis slider on the bottom that allows you to look at signal levels at each stage of the sequence.
       Change the following parameters and press "STOP" followed by "PUSH" each time and observe what changes *in the diagrams*:
         - Slice thickness

- Repetition time
- Flip angle
- Flow speed (hint: use the time axis slider)

b. Effects of flow speed
Make sure you are on the "BRIGHT BLOOD" tab. Set slice thickness = 5 mm, repetition time = 250 ms, echo time = 5 ms. Set flow speed to 5 mm/s and push the syringe. The diagram should be updated shortly after the flow begins.

Now, look at the slider on the bottom of the diagram. The time point marked as "RF1" means the state of the spins right after the first RF pulse and the time point marked as "TR1" indicates the state after a time of TR passes following the first RF pulse (right before the second RF pulse).
- Drag the slider to "RF1". We see that the initial "0" state is replaced by a "1" state after the RF pulse, which means that the entire slice (orange) has experienced one RF pulse.
- Drag the slider to "TR1". Because time has passed, there is new inflow ("fresh blood") from the left to the right across the slice. The inflow occupies ¼ of the slice because the net movement is (5 mm / s) (250 ms) = 1.25 mm while the slice thickness is 5 mm. This new flow was outside the slice, so it has experienced 0 RF pulse.
- Drag the slider to "RF2". A second RF pulse has been applied, so both parts of the slice get +1 pulses applied.
- Drag the slider to "TR2" and then "RF3" The same effect as in "TR1" happens and the same amount of new spins has entered the slice while the previous segments keep accumulating pulses.
- Eventually, this process will divide the slice into four segments experiencing 1, 2, 3, and 4 RF pulses, respectively. The brightness of the segments corresponds to their signal level and *fewer RF pulses generate more signal (explained later)*, except for 0 pulses, which means no signal.

Change flow speed and observe the effects on the diagram: look for the number of cycles needed to reach a "steady state" (where 2 states repeat themselves) and the number of segments.
- Does a faster flow generate more or less signal than a slower flow?

After each change of parameters on the left, press "transfer parameters" and then "RUN" to acquire an image. You may display the image as "Full range" (black = 0% signal and white = 100% signal) or "Normalized" (black = minimum signal present and white = maximum signal present). Hover over the image to read the actual signal levels in different regions: flow (the two circles), main cylinder (the large circle), and background (outside the circle).

   c. Effects of slice thickness
      Change slice thickness and observe the effects on the diagram and the image.
      - Does a thicker slice increase or decrease the flow signal? Note: the final signal is equal to the average signal across the slice.
   d. Effects of TR
      Change TR and observe the effects on the diagram and the image.
      - What does TR signify in the sequence (top) diagram?
      - How does TR change the number of slice segments? Why?
      - How does TR change the signal at different numbers of RF pulses?
   e. Effects of flip angle
      Change the flip angle and observe the effects on the diagram and the image.
      - How does the flip angle affect the top diagram?
      - What happens at flip angle = 90 degrees? How is it different from, say, flip angle = 15 degrees?
      - How does increasing the flip angle change the image?

3. Explore the spin echo dark blood effect
   a. Observe the spin echo diagram
      Go to the "DARK BLOOD" tab and press "PUSH". There are two plots: the top one showing RF pulses (a 90-degree pulse at time = 0 and a 180-degree pulse at time = 0.5 x echo time) and the magnetization, and the bottom one showing flow. There is a time axis slider on the bottom that allows you to look at signal levels at each stage of the sequence.

      Change the following parameters and press "STOP" followed by "PUSH" each time and observe what changes *in the diagrams*:
      ● Slice thickness
      ● Echo time
      ● Flow speed (hint: use the time axis slider)

   b. Effects of flow speed

Make sure you are on the "DARK BLOOD" tab. Set slice thickness = 5 mm, echo time = 50 ms. Set flow speed to 20 mm/s and push the syringe. The diagram should be updated shortly after the flow begins.
- The time slider should be at "0-" and there is no signal on the bottom flow diagram. Drag it to "0+". The portion within the slice (orange) lights up and indicates which spins experienced a 90-degree pulse at time = 0. This corresponds to the red vertical line on the pulse sequence diagram and the signal level is maximum as seen in both the "90 only" and the "180 only" curves that intersect at (time = 0, signal = 1).
- Go to time "(TE/2)-". This is right before the 180-degree pulse (orange vertical line) and we see that the signal has both moved to the right and decreased due to $T_2^*$ decay.
- Go to time "(TE/2)+". This is right after the 180-degree pulse. Now, while the same slice has been excited, it is divided into two parts: the inflow has only experienced the 180-degree pulse, while the previous portion now has experienced both pulses. The 180-degree pulse alone generates no signal in the fresh inflow and does not change the signal in the previously excited spins.
- Go to time "TE". Due to the Spin Echo effect (explained in definitions), the 90+180 portion lights up (topmost circle at time = TE on the top diagram). The 90-only portion has further decayed and shows a low signal (middle circle) and the 180-only portion shows no signal (bottom circle). You can hover over these circles to read the signal levels (the display label shows the (time, signal) pair).

Experiment with the flow speed and see what changes in the diagram and what doesn't.
- How does the flow speed affect the relative portions of "180 only", "90 + 180" and "90 only"?
- For each flow speed, press "TRANSFER PARAMETERS" and then "RUN" to acquire an image (make sure the plunger is moving when you acquire). How would you explain this change in contrast between the flow and the background?
- What happens to the image eventually if you keep increasing the flow speed? Why? (hint: observe all time points on the bottom diagram at a high flow speed)

c. Effects of slice thickness

Change the slice thickness and observe what happens to the diagram and the image after transferring parameters.
- What happens to the contrast as you change the thickness? Why? Hint: the final signal is proportional to the sum of all signals divided by the slice thickness (the average).
- What eventually happens if you keep decreasing the slice thickness? Explain this with the diagram.

d. Effects of echo time (TE)
Change the echo time and observe what happens to the diagram and the image.
- How does the flow speed affect the relative portions of "180 only", "90 + 180" and "90 only"? Hint: the amount of inflow is equal to (flow speed) x (0.5*TE).
- How does the contrast change as you increase the echo time?
- What eventually happens if you keep increasing the echo time? Explain this with the diagram.

4. (Optional) Explore the effects of tissue property
   a. The effects of $T_1$ on bright blood imaging
   Select the "BRIGHT BLOOD" tab. Set flow speed to 5 mm/s, slice thickness to 5 mm, repetition time to 250 ms, and echo time to 5 ms. Push the plunger and cycle through the diagram. Acquire an image with the same parameters.
   - Press "TISSUE PROPERTIES" and change $T_1$ from 2000 ms to 500 ms. What changes in the diagram and in the image?
   - How does the flow signal depend on $T_1$?
   - How does the background signal depend on $T_1$? Hint: try setting the flow speed to a level so that more than 7 or 8 segments are created and observe the behavior of the signal from spins that receive many RF excitations.
   b. The effects of $T_2$ on bright blood imaging
   On the "BRIGHT BLOOD" tab, change $T_2$ and generate a new diagram and acquire an image each time. The gray line in the top diagram indicates where the signal is received and the small green circle indicates the signal level The signal level depends on the $M_{xy}$ line because any magnetization needs to be rotated into the xy-plane to be detected.
   - How does $T_2$ change the $M_{xy}$ line?
   - How does $T_2$ affect the image?

- Change the echo time and explain how it affects bright blood imaging.

c. The effects of $T_1$ on dark blood imaging
Select the "DARK BLOOD" tab. Change $T_1$ and observe the diagram and images. You will learn more in Game 6 but $T_1$ determines how much $M_z$ recovers over a given amount of time that is free of RF pulses.
- How does $T_1$ affect the diagram? Why?
- How does $T_1$ affect the image? Why? Hint: for the Spin Echo Dark Blood sequence, TR is made quite long so each pair of 90-180 pulses is played a long time after the previous pair.

d. The effects of $T_2$ on dark blood imaging
Select the "DARK BLOOD" tab. Change $T_2$ and press the "PUSH" button each time.
- How does $T_2$ affect the top diagram? As you will learn in Game 6, $T_2$ determines how fast $M_{xy}$ decays over time.
- How does $T_2$ affect the bottom diagram?
- How does $T_2$ affect the image? Why?

## Questions

1. How can we visualize blood vessels with MRI?
a. By eliminating signal from background tissues
b. By eliminating signal from blood
c. By eliminating signal from blood vessel walls
d. By performing a maximum intensity projection on any MR image

2. Which of the following is false about steady state SPGR?
a. The steady state signal is lower than the initial signal
b. The fewer RF pulses protons experience, the higher signal they generate
c. The steady state signal is dependent on the flip angle
d. Fewer RF pulses are needed for moving protons to reach steady state than for stationary protons

3. How would you describe the RF pulse pattern in SPGR?
a. A train of equally spaced, alpha-degree (alpha < 90) pulses
b. A train of equally spaced 18 0-degree pulses
c. A 90-degree pulse followed by equally spaced 180-degree pulses
d. An alpha-degree (alpha < 90) pulse followed by equally-spaced 90-degree pulses

4. Which of the following is true about T2 and T2*?
   a. Spin echoes generate T2* contrast
   b. T2* is always longer than T2
   c. T2 and T2* are the same for moving spins
   d. T2*-weighted signal is lower than T2-weighted signal

5. How do you describe the RF pulse pattern in SE?
   a. 90-deg pulse, Δt, 180-deg-pulse, Δt, get signal
   b. 180-deg pulse, $Δt_1$, 90-deg pulse, $Δt_2$, 180-deg pulse, $Δt_2$, get signal
   c. 90-deg pulse, 2Δt, 180-deg pulse, get signal
   d. 90-deg pulse, Δt, 180-deg-pulse, 2Δt, get signal

6. Which of the following is false about maximum intensity projection?
   a. It generates a 2D image from a 3D volume
   b. High signals are emphasized
   c. You should use it for dark-blood MRA images
   d. It helps radiologists visualize the shape of a vascular tree

# Game 5: Proton's got moves

## Why?
Nuclear magnetic resonance is the key physical phenomenon and the basis for MRI. The nuclei of hydrogen atoms everywhere in our bodies always spin around its axis, generating tiny magnetic moments. To interact with them, we must talk in their magnetic language by applying 3 different types of magnetic fields at appropriate frequencies so they dance in harmony and emit a signal we can listen to. If you understand the cool moves of protons, you can be a spin choreographer and make them reveal what's inside you.

## Materials
- Gyroscope
- Copper coil
- Small bar magnet
- Multimeter

## Background
1. Key terms
   a. **Spin**: Protons, neutrons, and electrons all have **intrinsic** angular momentum, also called spin. These subatomic particles are always rotating around their own center axes by nature. In the image, the Angular Momentum vector (L) of a cylinder points along the right-handed axis of its rotation.

   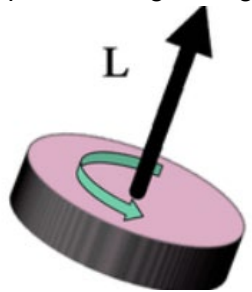

   Spin - Questions and Answers in MRI (mriquestions.com)

   b. **Magnetic moment** is a vector that measures how much of a magnet a thing is and the direction of this magneticness. A loop of wire carrying a current has a magnetic moment, and so does a fridge magnet. Protons are magnets too, and each of them has a fixed magnetic moment μ.

   c. **Net Magnetization (M)**: Because the protons and their magnetic moments are small, it is preferable to think about the sum of their behavior rather than trying to model each individual proton. Net Magnetization (M) is the average vector of all the proton magnetic moments within a volume in space.

When there is no outside magnetic field, the spins are all pointing at random directions and M is zero. When you turn on such a field and keep it on, however, net magnetization develops and ends up larger for higher magnetic field strengths because of energy level effects.

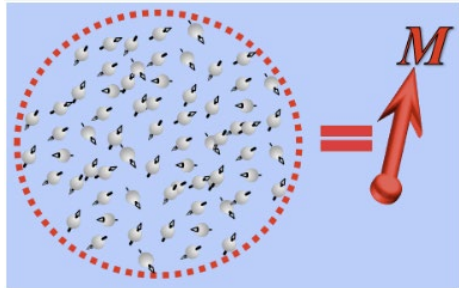

Spin - Questions and Answers in MRI (mriquestions.com)

d. **Main magnetic field** is a strong magnetic field (clinically, about 30000 to 70000 times Earth's field) essential to MR. It is made highly uniform and points along the z axis. When you go into the MR scanner, you are under its influence and a net magnetization vector develops in you.

e. **Precession** is the motion of the protons as they spin around the axis of the main magnetic field in a cone like manner. This happens partly because the protons have angular momentum, and partly because the main magnetic field is pulling on the proton's magnetic moment.

You can see similar effects with spinning tops when they go off axis and rotating wheels suspended on strings at an angle. Precession gets faster when you use stronger magnetic fields.

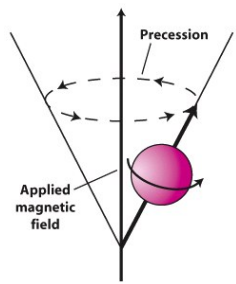

https://sites.duke.edu/apep/module-3-alcohol-cell-suicide-and-the-adolescent-brain/explore-more/mri-the-dance-of-the-whirling-protons/

f. **Rotating frame of reference:** Since the magnetic moment is rotating so fast, its movement can quickly become very complicated when we start turning on additional magnetic fields. The way to simplify this motion is to pretend that we are also going around the precession axis at exactly the same speed, in a mad merry-go-round kind of way. Once we are on this bandwagon, the spin's magnetic moment seems to stay still. It is just like how two people standing on two trains side by side moving at the same speed seem stationary to each other.

g. **Radiofrequency (RF) pulses** are how we talk to spins using a second, more short-lived, magnetic field. The RF field is perpendicular to the main field and has to rotate as fast as the spin's precession to catch up with it. Once it catches up, it flips the spins as if turning a wrench, knocking them out of the comfortable z equilibrium and towards the xy plane. Then, after we've rotated the desired amount, the pulse is turned off.

The axis of rotation is along the RF field vector itself and the angle is proportional to both the strength and the duration of the pulse. The figure below shows how M is turned by 90 degrees in (b) and then spreads out because individual spins have slightly different frequencies in real life.

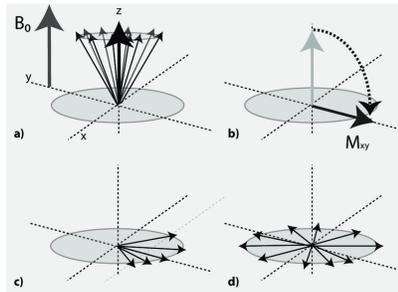

https://www.researchgate.net/figure/e-application-of-an-RF-pulse-and-T2-relaxation-After-an-RF-pulse-has-flipped-all_fig5_42788125

h. **Nutation**: It is just another word for RF rotation of the net magnetization vector to distinguish it from precession.

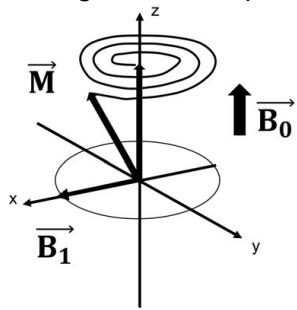

https://www.researchgate.net/figure/Nutation-at-Larmor-frequency-of-the-magnetization-vector-around-the-RF-pulse-B-1-applied_fig2_348670030

i. **Electromotive force**: This is the electrical potential difference, or voltage, created, when the magnetic flux across a looped wire, also called a coil, changes. This happens only when a coil is placed correctly, in a way that magnetic fields generated by the net magnetization M pass through it. When M is in precession, a sinusoidal voltage can be measured across the coil.

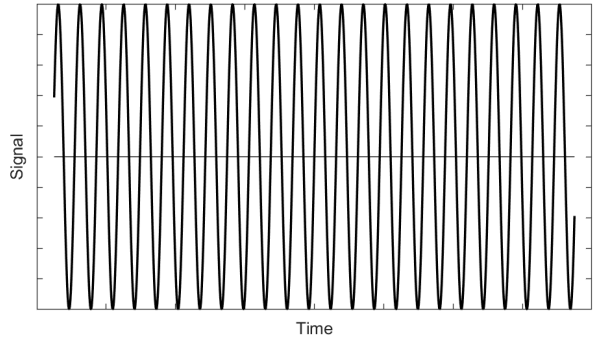

j. **Free induction decay**: This is the more realistic emf signal that occurs right after the 90 degree RF pulse. Due to interactions between spins causing them to have slightly different frequencies over time, the signal goes down to zero exponentially at predictable rates described by the T2 tissue parameter.

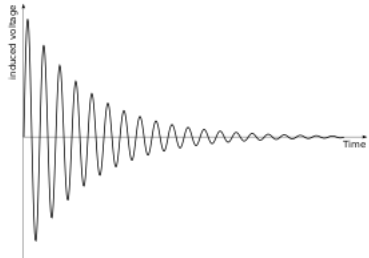

https://en.wikipedia.org/wiki/Free_induction_decay

2. Lab procedures:
    a. "The Equilibrating Move" - Equilibrium Magnetization
        i. Turn on the main field by clicking on 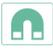. Once it is turned on, adjust the main field value and note down your observations.

        As the main field is turned up, ______________________________.

    b. "The Circulating Move" - Precession
    i. Under guidance of the mentor, take the gyroscope (or bicycle wheel, or spinning top) and tip it at an angle. Release. Then repeat the step after making it rotate around its own axis first. What are the forces acting on the gyroscope and how does its motion change when it's being spinned versus when it's not?

    ii. Make sure the rotating frame is turned off. Then set the initial magnetization to have a nonzero x or y component (for example: theta = 45, phi = 45, size = 1).

iii. Hit the "RUN" button. What do you observe? How does each component of the magnetization vector (Mx, My, Mz) change in time?

- Mx :
- My :
- Mz :

iv. Change the main field strength and hit "RUN" again. What is the difference?

v. Turn on the rotating frame. We are now essentially traveling on a merry-go-round that's matched to the spin's frequency. What do you see now?

vi. Imagine for a moment and answer: what do we see if our merry-go-round is going faster than the spin? What if we are going slower?

c. "The Tipping Move" - RF pulses
   I. Push "RESET" to turn off everything. Then turn on B0 and wait until the equilibrium M is fully developed.
   ii. Push the button "Show RF" to display the direction of the RF magnetic field.
   iii. Adjust the flip angle and pulse direction
   iv. Make sure both the main field and the rotating frame are turned on. Then push "Tip". The spin should move.
   v. Repeat i-iv with five different values and note your observations in the table below:

| Flip Angle | Pulse Direction | Observations |
| --- | --- | --- |
|  |  |  |
|  |  |  |
|  |  |  |
|  |  |  |

Vi. On the "set initial magnetization" panel, you can choose to have M0 start with any direction. Try to make the spin vector rotate to the designated final positions for the following settings with the RF pulse and record the steps. You may use up to 3 RF pulses (FA: flip angle; DIR: pulse direction). The fewer pulses you use, the more kudos you get!

| Initial M (always use \|M/M0\| = 1) | Final (Mx,My,Mz) | FA 1 | DIR 1 | FA 2 | DIR 2 | FA 3 | DIR 3 |
|---|---|---|---|---|---|---|---|
| theta = 90, phi = 90 | (0,0,1) | | | | | | |
| theta = 90, phi = 0 | (-1,0,0) | | | | | | |
| theta = 90, phi = 0 | (0,1,0) | | | | | | |
| theta = 45, phi = 90 | (1,0,0) | | | | | | |
| theta = 0, phi = 0 | (0,1/2,-sqrt(3)/2) | | | | | | |

d. "The Electrifying Move" - Signal induction

I. Connect the wire loop to the spectrometer and try moving the bar magnet in different ways: rotating next to the coil, moving towards and away from the coil, moving parallel to the coil, and so on. How can you generate the most signal?

ii. Push the "RESET" button to turn off everything. Then turn on B0 and wait until the equilibrium M is fully developed.

iii. Push the bull's eye icon to turn on the receive coil.

iv. Is there a signal now? Why?

v. Set the magnetization to x or y. Then hit "RUN". Is there a signal now? Why?

vi. Change field strengths and hit "RUN" again each time. Record your observations below:

| Field strength (gauss) | Effects on spin | Effects on signal plot |
|---|---|---|
| | | |
| | | |
| | | |
| | | |
| | | |

vi. Set the initial magnetization to have different values of theta and/or phi and hit "RUN". Record your observations.

| Theta | Phi | Observations |
|-------|-----|--------------|
|       |     |              |
|       |     |              |
|       |     |              |
|       |     |              |

What matters in determining signal amplitude? What does not?

3. Questions

(1) What is the difference between a 90 degree RF pulse and a 180 degree RF pulse?
   a. The 90 degree pulse is shorter than the 180 degree pulse
   b. The 90 degree pulse is at a right angle to the spin magnetic moment while the 180 degree pulse is antiparallel to it
   c. The 90 degree pulse rotate half the number of spins compared to the 180 degree pulse
   d. The 90 degree pulse tips the equilibrium M to the x-y plane while the 180 degree pulse does not

(2) Why do we need to apply an RF pulse?
   a. The spins absorbs the RF energy and converts it into heat, which can then be detected by the coil
   b. The RF pulse allows spins to develop a magnetic moment which is necessary for signal generation
   c. The spin magnetic moments start out along z but cannot be detected unless tipped so they lie in the x-y plane
   d. The spins precess around the transmitted RF field to generate emf in the receiving coil

(3) What happens to the emf signal when the main magnetic field is turned up?
   a. The signal oscillates faster
   b. The signal oscillates more slowly
   c. The signal decays faster
   d. The signal decays more slowly

(4) Among the four options, which one maximizes the voltage range of the received signal at a constant main field? The initial magnetization is at equilibrium and points along z.
   a. Use a 5-degree pulse with a phase of 180 degrees
   b. Use a 85-degree pulse with a phase of 45 degrees
   c. Use a 179-degree pulse with a phase of 0 degrees
   d. Use a 225-degree pulse with a phase 0f 76 degrees

# Game 6: Relaxation Station

**Why?**
As we saw in Game 3, parameters like TR and TE change the contrast of our image the way different color filters create effects on digital photos. As you vary these imaging parameters, the signal shifts for every pixel. However, this doesn't explain why the signal of a single image can be different between tissue types - why is it that gray matter looks darker than white matter on a "T1-weighted" image? What does "T1-weighted" even mean? In this game, you will learn what T1 and T2 mean and how to affect the signal evolution. These are two parameters with units of time (in, say, milliseconds) that differ between tissues and interact with the sequence settings to produce the final signal. A change in T1 or T2 may indicate that something is wrong in a given soft tissue and this is why we can see brain tumors with MRI.

**Materials**
- Water phantom
- Oil phantom
- Dosed phantoms (T1 series x 3 , T2 series x 3)

**Background**
1. Key terms
    - Longitudinal relaxation and T1
    - Transverse relaxation and T2
    - Curve fitting
    - Inversion Recovery Spin Echo (IRSE)
    - T1 mapping
    - Turbo Spin Echo (TSE)
    - T2 mapping
2. Basics
   In Game 5, we explored the four moves of the net magnetization, **M**: (1) equilibrium; (2) precession; (3) nutation; and (4) signal generation. The simulations for moves 2-4 then were all based on a set of equations (called the Bloch equations) that describe how M changes over time depending on which magnetic fields were applied:

$$\frac{dM(t)}{dt} = \gamma M(t) \times B(t)$$

   (M(t), B(t) are vector functions of time representing the magnetization and the magnetic field; γ is a constant called the gyromagnetic ratio with units of [radians/Tesla]).

   This equation leads to two facts when there is a constant $B_0$ field along z (as it is for MRI): first, a M vector initially at an angle to the z-axis will rotate around z (precession) indefinitely and its size will not be diminished; second, a M vector initially aligned with z

Just like there is no 100% efficient machine in reality, we encounter signal losses from T1 and T2 relaxation processes which are included in the "Bloch equations with relaxation":

$$\frac{dM(t)}{dt} = \gamma M(t) \times B(t) - \left(\frac{M_x(t)}{T_2}, \frac{M_y(t)}{T_2}, \frac{M_z(t)-M_0}{T_1}\right)^T$$

Don't worry about the details of this vector differential equation; we will explain its effects on the magnetization's movement shortly. In general, including these terms make the M vector "go home" - that is, after the RF field is turned off, the spins tend to return to the equilibrium : $M_x = 0$, $M_y = 0$, and $M_z = M_0$. T1 and T2 determine how fast it "goes home".

3. Explanations
   <u>Longitudinal relaxation and T1</u>
   Longitudinal relaxation happens when individual spins drift randomly because of fluctuating local magnetic fields and gradually go back to the state with the least energy, which corresponds to $M_z = M_0$ when $B_0$ is oriented along +z. The same process happens when the initial magnetization develops in Move 1 of Game 5 after $B_0$ is turned on.

   Mathematically, the T1 term in the Bloch equation contributes a positive term to the derivative of $M_z$ since it is always true that $M_z <= M_0$. What results is the following behavior of $M_z$ that has been set to the specific value of $M_z(0)$ at t = 0:

   $$M_z(t) = M_0 - (M_0 - M_z(0))e^{-t/T_1}$$

   The difference between the initial $M_z$ and $M_0$ decays exponentially.
   The meaning of T1 can be found by setting t = T1: the exponential term becomes 1/e = 0.37, so T1 is the time during which the "distance to the goal" has been reduced to 37% of its full value.

   After a 90-degree pulse, $M_z(0) = 0$, and $M_z(t)$ can be simplified as:

   $$M_z(t) = M_0(1 - e^{-t/T_1})$$

   The plot below shows its evolution:

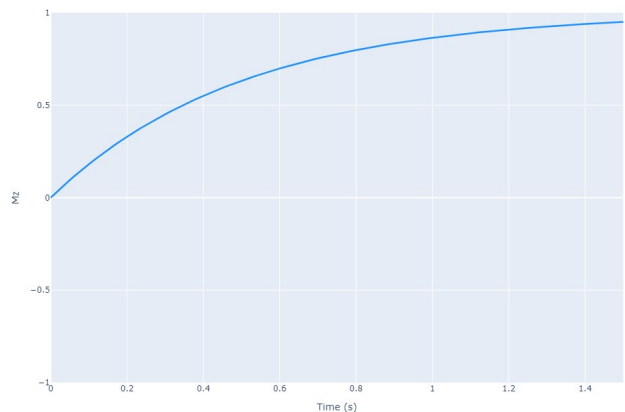

   After a 180-degree pulse, $M_z(0) = -M_0$, and $M_z(t)$ can be simplified as:

   $$M_z(t) = M_0(1 - 2e^{-t/T_1})$$

The plot below shows its evolution:

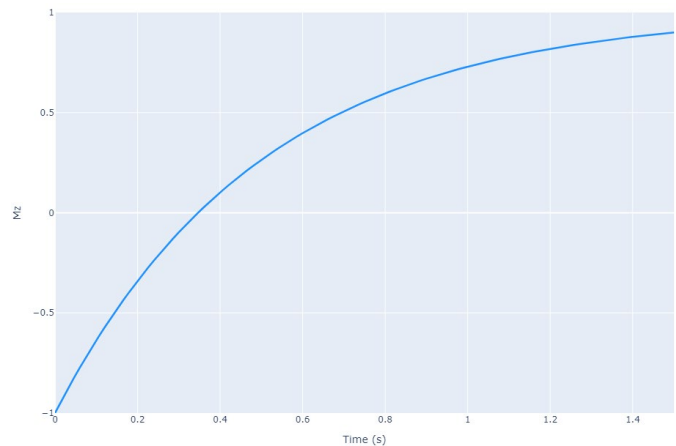

$T_1$ is different between tissues and changes with the main field strength. At $B_0=3T$, some approximate values in the brain are:

> Gray matter: 1600 ms
> White matter: 840 ms
> Cerebrospinal fluid: 4000 ms

You will be able to see what different T1 curves look like in the simulation part of the game.

Transverse relaxation and T2
After the 90-degree pulse flipped the magnetization vector into the x-y plane, the vectors started rotating around the z-axis at slightly different speeds. Some get ahead and others lag behind. Because the signal is equal to the vector addition of all these spins, the signal decays! One example is given here. The spins dephase at a rate of 30 degrees per millisecond. At t = 0, the signal along x is is $3M_0$; at t = 1 ms, the slower and faster spins as both rotated 30 degrees (up and down, respectively) and the signal is reduced to $2M_0$ - this is called dephasing as the spins develop different angles, also called their phases; at t = 3 ms, the rotation angle becomes 90 degrees and only the center spin has any component along x, so the signal is $M_0$. A plot of the signal proportions shows that a non-linear decay happens. There would be a spread of spins instead of just three with different probabilities in real life, but the same effect of dephasing happens.

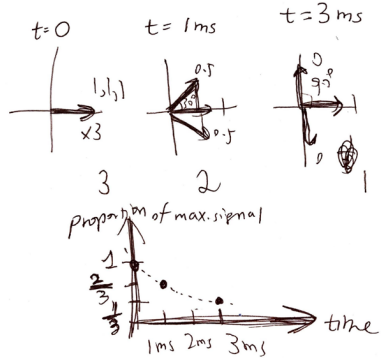

The more realistic model can be derived from the "Bloch equations with relaxation" and is given by:

$$M_{xy}(t) = M_{xy}(0)e^{-t/T_2}$$

This curve is shown in a plot below. It is a simple exponential decay from the initial $M_{xy}$, also called the transverse magnetization (as opposed to the longitudinal magnetization, $M_z$). Here we define $M_{xy}$ to be the length ("absolute value") of the 2D vector ($M_x$, $M_y$).

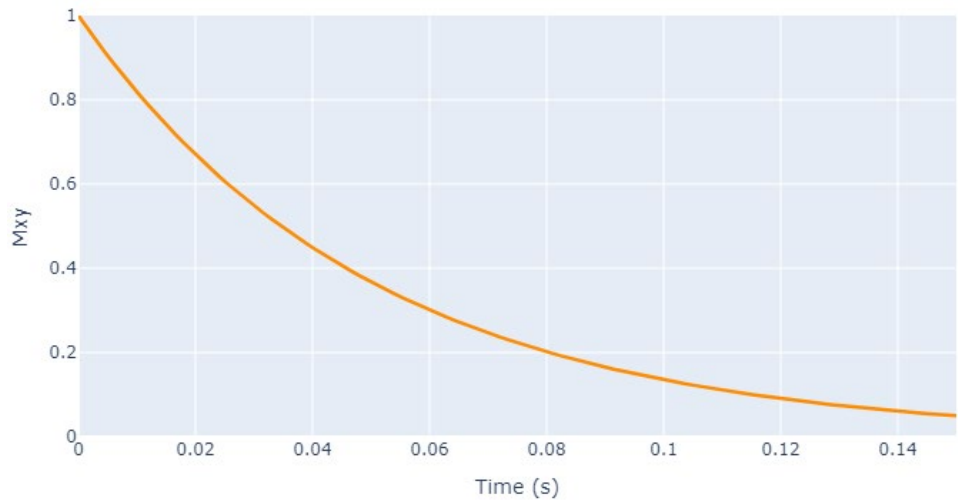

At t = T2, $M_{xy}$ has been reduced to 37% of its initial value.
At $B_0$ = 3T, some approximate T2 values in the brain are:

> Gray matter: 90 ms
> White matter: 70 ms
> Cerebrospinal fluid: 2000 ms

You will be able to see what different T2 curves look like in the simulation part of the game.
Curve Fitting

Curve fitting is a procedure for fitting paired data $\{(X_i, Y_i)\}$ with a predefined model. For example, we can fit the fairly linear data below (red) with the linear model of $y = ax + b$ (black):

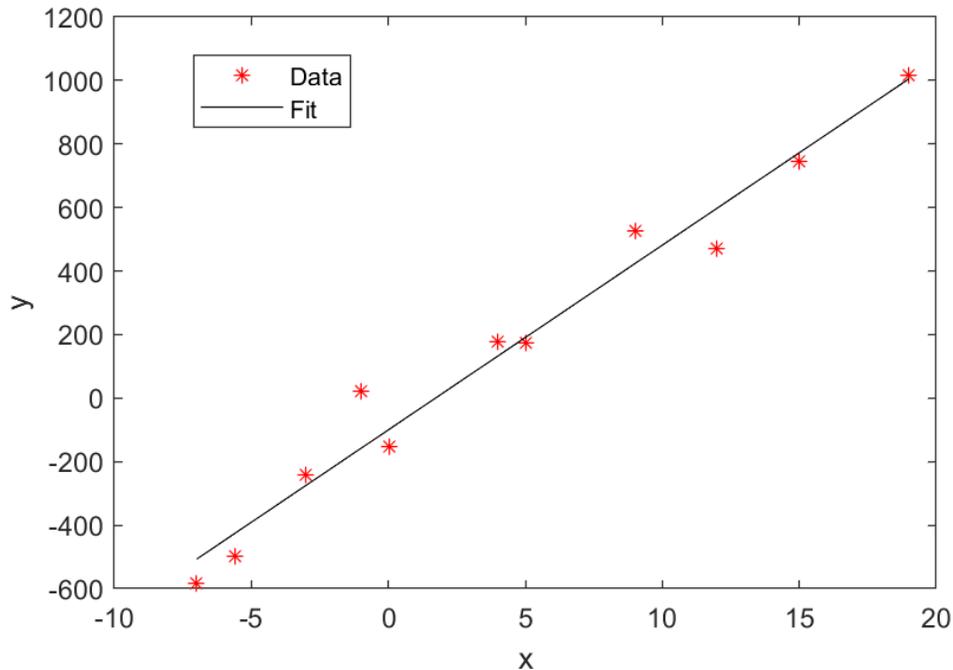

The fit is chosen so that the sum of squared vertical distances between the data (y) and the regression line is minimized and is an estimate of the relationship between x and y. For example, the estimated slope indicates how much y changes with a unit change in x. A similar procedure can be performed for non-linear models.

Inversion Recovery Spin Echo (IRSE)
IRSE is an MRI sequence with the following steps:
1. Apply a 180-degree pulse so M points along -z;
2. Wait for a time of TI ("inversion time");
3. Apply a 90-degree pulse so the remaining $M_z$ at that point gets converted into $M_{xy}$
4. Acquire data at TE ("echo time")

During the waiting period of TI, $M_z$ grows towards $+M_0$. How much it grows depends on both the tissue parameter T1 and the sequence parameter TI. The final signal is given by:

$$s(TI) = C|1 - 2e^{-TI/T_1}|e^{-TE/T_2}$$

where C is a scaling constant.

T1 mapping

We can now apply the IRSE sequence multiple times, each time with a different TI. Combining information across TIs will allow us to map out, voxel by voxel, the T1 values across the image. This is done with the following steps:

1. Acquire IRSE at different TIs to get a series of images;
2. For each voxel on the image, plot its signal against TI.
3. Perform a curve fitting with the T1 model for each voxel. This produces a T1 value as one of the fit parameters.
4. Display the T1 values as a map.

This map may serve as a quantitative tracker of disease states.

Turbo Spin Echo (TSE)
TSE is an MRI sequence with the following steps:
1. Apply a 90-degree pulse
2. Wait for a time of Δt
3. Apply a 180-degree pulse
4. Acquire data from the Spin Echo (see Game 4) at t = 2Δt = $TE_1$
5. Repeat steps 2-4 multiple times to generate further spin echoes at increasingly large TEs until signal dies off.

This allows us to sample different points on the T2 decay curve efficiently: {$TE_n$=2nΔt} are sampled for n = 1, 2, 3, …, N where N is the total number of spin echoes generated. The signal depends on TE and is given by:
$$s(TE) = Ce^{-TE/T_2}$$
where C is a scaling constant.

T2 mapping
We can now apply the TSE sequence multiple times and get multiple TEs in one go. Enough spatial information is collected so that one image can be reconstructed for each TE. Combining information across TEs will allow us to map out, voxel by voxel, the T2 values across the image. This is done with the following steps:

1. Acquire TSE at multiple TEs to get a series of images;
2. For each voxel on the image, plot its signal against TE.
3. Perform a curve fitting with the T2 model for each voxel. This produces a T2 value as one of the fit parameters.
4. Display the T2 values as a map.

This map may serve as a quantitative tracker of disease states.

**Lab procedures**
1. Explore T1 simulation
   a. Interact with the T1 spin animation
   Select "T1" among the left tabs and press "Simulation". Bring your attention to the leftmost of the three plots. You can press "Play" inside the plot to see how the blue M vector changes in time. Experiment with different values of "Init Mz (% M0)", pressing "Simulation" each time, and describe how it affects the spin animation.
   b. Inspect the T1 relaxation curve
   Look now at the middle plot, which shows $M_z$ as a function of time. Change "Initial $M_z$" to -1, -0.5, 0, 0.5, and 1, and observe how it modifies the curve. Does this agree with your findings in (a)? Next, set "Initial $M_z$" to 0 and change T1 to 100 ms, 500 ms, 1000 ms, and 4000 ms in turn. Hover your mouse over the curve; what is the signal level at t = T1 for each of the values (you can change the "Duration" to simulate longer or shorter periods of time)? What does this tell you about the meaning of T1 as a time constant?
   c. Plot the T1 mapping sequence: IRSE
   The right plot shows the IRSE sequence. Change the Inversion Time to 10 ms, 50 ms, 150 ms, and 600 ms and observe changes in the $M_z$, $M_x$ curves. When does the second pulse happen, and how does it relate the $M_z$ and $M_x$ curves? Can you choose an Inversion time so that the $M_x$ signal is minimized? Note: the final signal detected is proportional to the $M_x$ value right after the second pulse (90-degree).

2. Explore T1 mapping
   a. Acquire T1 images
   Select "T1" among the left tabs and press "Mapping". Press "SCAN" to acquire images at each TI listed in the "TI array (ms)" field. The left plot now shows a series of images at those TIs. The four circles are solutions with different T1 values. Cycle through the images with the slider. How does signal shift for each of the spheres as TI increases? You can use the fields above "SCAN" to customize your TIs before reacquiring.
   b. Perform region-of-interest curve fitting
   Select one of spheres 1, 2, 3, or 4 above "FIT". The sphere selected will be displayed on top of the images as a green circle and the average signals in that sphere (here, it is our region-of-interest or ROI) will be plotted. Do the signals correspond to what you see on the left? Note that the images show absolute values of the signals.
   Now, you can perform a curve fitting by pressing "FIT". How does the curve change between the four spheres? Does the curve describe the signals perfectly?
   c. Generate a T1 map!

Press "MAP" to automatically perform curve fitting on all voxels. This might take a bit of time. While waiting, feel free to select the "T1 phantom" and look at the true T1 values of the model. When the map is done, it will be displayed. How does the map differ from the true T1 values? What caused this?
   d. (Optional) Explore the effects of prescribed TIs on map quality
   Try using more TIs, fewer TIs, shorter TIs, or longer TIs to generate T1 maps and compare their qualities. Does this performance change between spheres?

3. Explore T2 simulation
   a. Interact with the T2 spin animation
   Select "T2" among the left tabs and press "Simulation". Look at the leftmost of the three plots. Press "Play" to start the spin animation. The orange line represents the x-component of the magnetization vector. Experiment with different values of "Init Mx (% M0)", pressing "Simulation" each time, and describe how it affects the spin animation.
   b. Inspect the T2 relaxation curve
   Look now at the middle plot, which shows $M_z$ as a function of time. Change "Initial $M_x$" to 0, 0.5, and 1, and observe how it modifies the curve. Does this agree with your findings in (a)? Next, set "Initial $M_x$" to 1 and change T2 to 5 ms, 50 ms, 100 ms, and 500 ms in turn. Hover your mouse over the curve each time; what is the signal level at t = T2 for each of the values (you can change the "Duration" to simulate longer or shorter periods of time)? What does this tell you about the meaning of T2 as a time constant?
   c. Plot the T2 mapping sequence: SE
   The TSE sequence is efficient at getting multiple TEs in one go, but the regular Spin Echo (SE) sequence as described in Game 4 can also be repeated at different TEs to produce the mapping data. The right plot shows the SE sequence. Change the Echo Time (TE) to 5 ms, 10 ms, 50 ms, and 100 ms and observe changes in the $M_z$ and $M_x$ curves. How does TE affect the signal level?

4. Explore T2 mapping
   a. Acquire T2 images
   Select "T2" among the left tabs and press "Mapping".. Press "SCAN" to acquire images at each TE listed in the "TE array (ms)" field. The left plot now shows a series of images at those TEs. The four circles are solutions with different T2 values. Cycle through the images with the slider. How does the signal shift for each of the spheres as TE increases? You can use the fields above "SCAN" to customize your TEs before reacquiring.
   b. Perform region-of-interest curve fitting
   Select one of spheres 1, 2, 3, or 4 above "FIT". Do the signals correspond to what you see on the left? Note that the images show absolute values of the signals. Now, you can perform a curve fitting by pressing "FIT". How does the curve change between the four spheres? Does the curve describe the signals perfectly?

c. Generate a T2 map!
Press "MAP" to automatically perform curve fitting on all voxels. This might take a bit of time. Select the "T2 phantom" and look at the true T2 values of the model. How does the map differ from the true T2 values? What caused this?

d. (Optional) Explore the effects of prescribed TEs on map quality
Try using more TEs, fewer TEs, shorter TEs, or longer TEs to generate T2 maps and compare their qualities. Does this performance change between spheres?

**Questions**
1. Which of the following is true about T1?
    a. T1 refers to transverse relaxation
    b. T1 relaxation causes $M_z$ to approach zero
    c. T1 relaxation causes $M_z$ to approach its equilibrium value
    d. T1 can be mapped by varying TEs

2. Which of the following is false about T2?
    a. T2 decay is caused by dephasing of spins
    b. T2 decay happens in the x-y plane
    c. T2 refers to the time taken to decrease $M_{xy}$ by about 63%
    d. T2 has units of [radians/Tesla]

3. Which of the following describes an IRSE sequence?
    a. 90 RF, wait for T, 180 RF, wait for 2T, acquire
    b. 180 RF, wait for 5T, 90 RF, wait for T, acquire
    c. 90 RF, wait for T, 180 RF, wait for T, acquire, wait for T, 270 RF, wait for T, acquire
    d. 90 RF, wait for T, 90 RF, wait for T, acquire

4. Which of the following describes a TSE sequence?
    a. 90 RF, wait for T, 180 RF, wait for T, acquire, wait for T, 180 RF, wait for T, acquire
    b. 180 RF, wait for T, 90 RF, wait for T, 180 RF, acquire
    c. 180 RF, wait for T, 90 RF, wait for T, acquire
    d. 90 RF, wait for 2T, acquire

5. What does curve fitting accomplish?
    a. It converts one image into a parameter (T1 or T2) map.
    b. It fits the T1 or T2 signal model to the mapping data acquired at different timings.
    c. It draws a straight line through T1 or T2 mapping data for each voxel.
    d. It traces tissue boundaries to perform a segmentation based on T1 or T2 maps

6. How can we get data to perform T2 mapping?
    a. Apply an IRSE sequence with different TEs
    b. Apply a TSE sequence with different TEs
    c. Apply a TSE sequence with different TIs
    d. Apply an IRSE sequence with different TRs

7. What is $M_z$ at t = 100 ms after a 180-degree pulse has been applied at t = 0 on a tissue type with T1 = 300 ms?
    a. - 43.3% $M_0$
    b. +27.4% $M_0$
    c. 0
    d. +91.0% $M_0$

# Game 7: Puzzled by Projection I

**Why?**

Projection is key to medical imaging technologies like X-ray, CT, and MRI. Each projection is a point of view that gives us a "shadow" of a higher-dimensional object. These "shadows" contain partial information, which can be combined to recover the original object in all its glory - for example, the inner structures of you!

**Materials**

- 3 letters phantoms (N, Y, C)
- 9 of basic shape phantoms (triangle, square, star)
- 3 of set 1 projection phantoms (half circles with central slit)

**Key Terms**

1. 3D coordinates
2. 2D projection
3. 1D projection
4. Projection imaging

**Background**

1. **3D coordinates**

    Any point in 3D space can be located with a 3D vector (x,y,z), which denotes its position relative to a known origin point. A 3D model is defined by surfaces, which are defined by triangles, which are in turn defined by its three vertices, each of which has coordinates $(x_i, y_i, z_i)$.

2. **2D projections** are easy to visualize. Imagine a brick suspended in air in a room with a top light that emits parallel rays:

    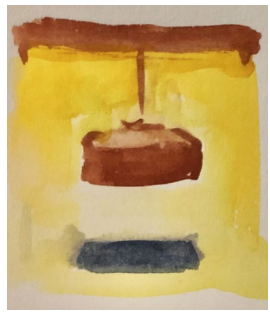

    The brick casts a rectangular, uniform shadow. Now, if we drill holes in the brick, they are going to pass light and change the shape of the shadow. If we start playing with different geometrical shapes, all of them will have shadows of different shapes and sizes.

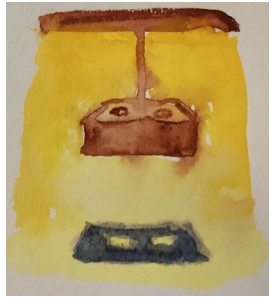

if we change the brick into a translucent block of glass, the shadow will be lighter, as more light comes through:

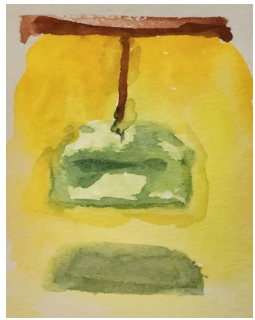

Real world objects (such as your brain) are made of different materials at different spots. Therefore, when light passes through them, the amount that reaches the other side varies and creates a weirdly shaded shadow. In this case, we are talking about quite high-energy light: X-rays! This is exactly how plain X-rays work!

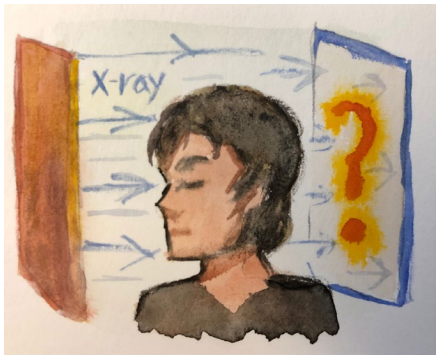

In this case, we see more light coming out the other side when there is less material for it to pass through. The amount of light can be measured on a light-sensitive sheet. Then the total amount of material the light had to pass through can be calculated for each point on the shadow. Brighter parts mean there was more material, and darker parts mean there was less. This is our 2D projection!

More mathematically, a 2D projection of a 3D function along an axis can be calculated by summing up the function $f(x,y,z)$ across all values on that axis. For example, a projection along the z axis can be calculated by creating a function $g(x,y)$, where:

$g(x0,y0)$ = (sum of $f(x0,y0,z)$ for all z between negative infinity and positive infinity)

If you've taken calculus, you know this can be done with an integral for continuous functions.

In the game, you'll be able to perform 2D projections along the x, y, and z axes.

3. **1D projections** turn 2D images into 1D curves just like 2D projections turn 3D volumes into 2D images. We can choose any angle from 0 degrees to 180 degrees to make the projection.

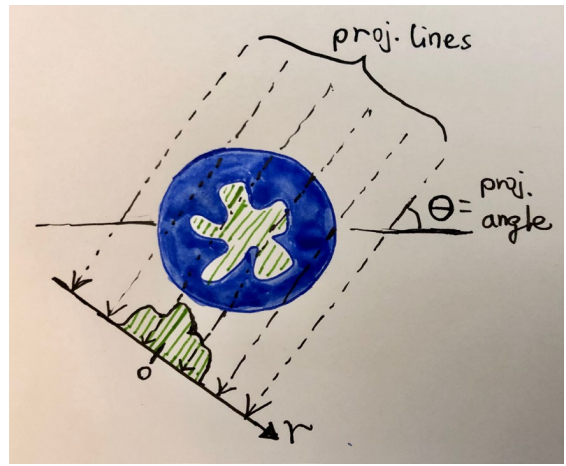

4. **Projection Imaging** works by making projections of the same subject and combining them to make images that show its internal structure. Examples include:
    - Plain radiograph or X-ray: uses X-rays to obtain a single 2D projection on X-ray-sensitive film or digital screen and views it directly as the image. In a chest X-ray, there is no front to back information and things overlap with each other.
    - Computed Tomography (CT): uses a rotating thin blade of X-ray to obtain many 2D-to-1D projections around a certain slice of your body (for example, around your waist or feet). These projections are combined, or <u>reconstructed</u>, to a single slice that shows clear internal structure and no overlap.
    - Projection reconstruction in MRI: certain types of MRI data can be reconstructed using CT-like methods. Why this is the case involves spatially varying magnetic fields and how hydrogen in your body responds to them as well as mathematical relationships between projections and spatial frequencies.

**Lab procedures**

1. Explore 3D models
    - Select one of the 3D models on the left panel and hit "load 3D model".
    - Rotate the model around and use the "transparent" button to observe its inner structures.
    - Repeat this with other models and view them from different angles.

        Can you know for sure about the model's 3D shape if you've only viewed it from one angle?

2. Explore 3D to 2D projection

- Select "z" for the projection axis and hit "show/hide lines" to display the direction of our 3D to 2D projection;
- Hit "2D projection" to display the projected image. What do you see?
- Make the 3D model transparent. Rotate the 3D model so that the projection lines are pointing out of the screen. Compare this view to the 2D projection.
- Repeat 1-3 for "x" and "y" for a few models.

In your words, what is 2D projection? If you can take one 2D projection of yourself, how would you do it and what information will the image tell you? This is exactly what X-ray machines do!

3. Explore 2D to 1D projection
   - Generate a 2D projection of the 3D model "N" in the "z" direction
   - Use the circle controller to select a projection angle of 90 degrees
   - Press "1D projection" to generate a 1D plot. You should see two bumps corresponding to the two legs of the letter "N".
   - Repeat with a projection angle of 0 degree. You should see a more flat curve. Because we are looking at "N" from the side, we can no longer tell the two legs apart.
   - Explore projection at various angles for the different phantoms.

   If we do enough 1D projections at various angles, we can figure out what the image looks like! This is how Computed Tomography or CAT scans work. Each new angle gives us a bit more information about the 2D slice. Some types of MRI also make images in this way.

4. Puzzle time!

   Now it's time for you to suss out some projections! Press "generate random model" to display some special cylinders and choose your favorite one. Don't generate the 2D and 1D projections until you have completed the table!

   Sketch these projections for the model:

   | Model name: _______________ | | | |
   |---|---|---|---|
   | 3D to 2D | Axis: x | Axis: y | Axis: z |
   | | | | |

| 2D to 1D | Angle: 90 degrees | Angle: 90 degrees | Angle: 90 degrees |
|---|---|---|---|
| | Angle: 0 degree | Angle: 0 degree | Angle: 0 degree |

Now you can press the projection buttons to check your answers. Do they make sense?

**Questions**

1. How many numbers are needed to locate a point in 3D space?
   A. 9
   B. 3
   C. 42
   D. 2
2. What of the following is true when you perform a 2D projection of a perfect 3D sphere?
   A. The projection will have the same gray value across the image
   B. The projection lines will bend to conform to the circular shape
   C. The projection will be the same from any angle
   D. The projection will be the same only when it's pointing along the x, y, or z axis.

3. What happens when you perform a 1D projection of a rectangle?
   A. The left, right, top, and bottom views are going to be the same
   B. The left/right views are the same, and the top/bottom views are the same
   C. The projection is always going to have right-angled corners
   D. The projection depends on how the rectangle is oriented in the x-y plane

# Puzzled by Projection II

**Why?**
In Game 7, we looked at how projections can capture one-sided views of an object and predicted projections given a model and an angle. Now, it is time to go at it the opposite way: to figure out what's inside an object by requesting projections of your choice! Indeed, this is essentially what we need to do when presented with raw MRI or CT data. In this game, you will be probing a mystery 3D object or 2D image with a limited number of 2D / 1D projections.

**Materials**
- Calibration tube
- Mystery tubes

**Background**
1. Key words
   - Forward and inverse processes
   - Encoding
   - Reconstruction

2. Basics
   This game will task you with figuring out the inner structure of a 3D model by performing and examining just a few projections. Are you up to the challenge?

3. Explanations
   Forward and inverse processes
   In daily life, we sometimes encounter pairs of *forward* and *inverse* processes. For example, a puzzle can be made by cutting a single picture into many pieces, which is a forward process. The re-assembly of the pieces would be the inverse process. Some processes are hard to invert: for example, a single 1D projection curve (explored in Game 7) cannot be used to determine for sure what the original 2D image was. However, if we combine 1D projections from many angles (like individual puzzle pieces!), we might be able to recover the image. In MRI, this inverse process is called reconstruction, whereas the initial forward process is called encoding. MR imaging involves many forward processes that need to be inverted so we can see the whole picture. We may not be able to perform the inversion perfectly but the goal is to have a good-enough estimate of the original image.

   Encoding
   We always encode information about space in MRI. That is, we need to tell where in 2D or 3D space a signal belongs so we can make an image by assigning a signal to each position. There are two ways of encoding space: Cartesian and non-Cartesian. In MRI, spatial encoding is the same as sampling from k-space (see Game 2), which has the same size as the image. Cartesian encoding makes sure we can sample k-space on a grid (think of a dotted notebook or wire crossings of a cooling rack, left figure). Non-cartesian is more general and includes all types of sampling that do not fall on a grid (right figure).

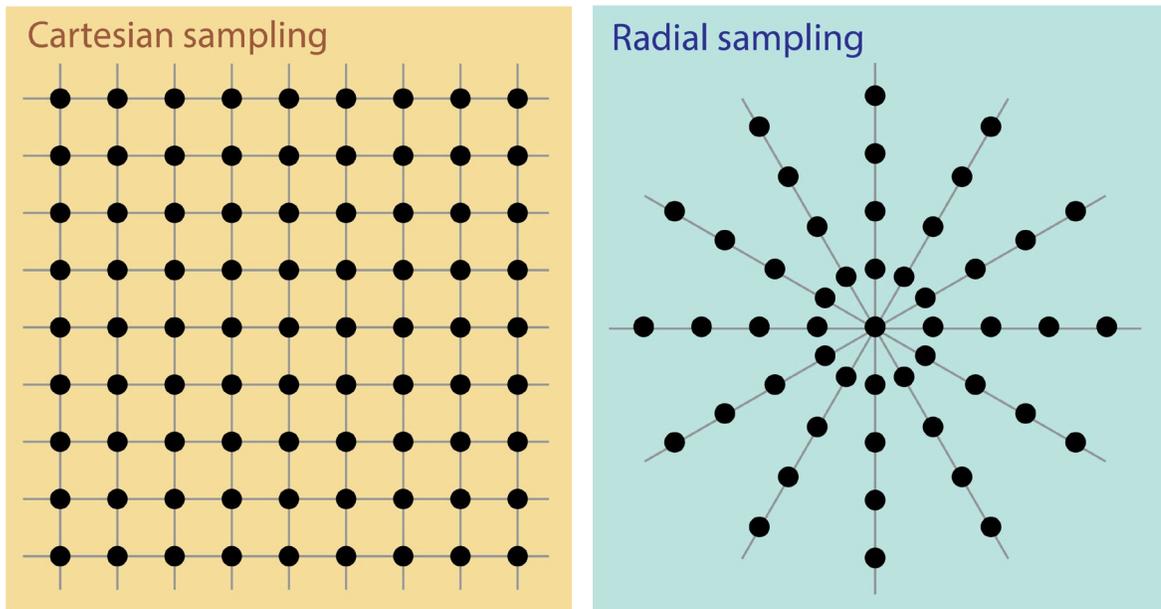

Reconstruction
- Reconstruction from Cartesian data is easy: take an Inverse Fast Fourier Transform (IFFT, the inverse process in Game 2), which can be done easily with code.
- Reconstruction from non-Cartesian data is trickier. There are many methods. Two of them are:
    - Backprojection: this only works for radially sampled data and is done in two steps: (1) Take a 1D Fourier transform of each radial line, which gives you *a 1D projection in the corresponding direction* (thanks to a mathematical fact called the Projection Slice theorem); (2) Perform a procedure called backprojection with this set of 1D spatial projections to get back the image. The figure below shows how backprojection works:

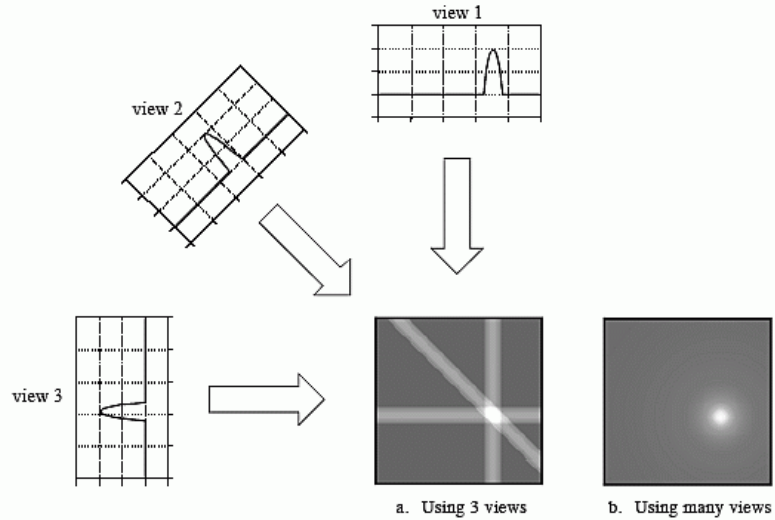

a. Using 3 views    b. Using many views

(https://www.dspguide.com/ch25/5.htm)
What we do is to take each projection and smear it across the image in the correct angle. The intersection of the projections highlights the location of the object and gets more accurate as more projections are included. Backprojection is also the basis for reconstructing CT (CAT scans) images!
- Gridding: First, we use the non-cartesian data to make an estimate of the cartesian data. Then we use the regular IFFT on the estimated cartesian data to reconstruct the image.

**Lab procedures**
In this game, instead of going over the math of these methods in detail, we would like to hone your intuition by asking you to "reconstruct" these images/volumes **just with your brain**. This way, you can appreciate that all the information we need to *make an image* or *tell several images apart* may be contained in one or more of its projections.

1. 2D puzzle
    a. Select "2D" on the left panel to enter 2D puzzle mode. A mystery image shows up! The "options" panel on the right shows multiple possible images to choose from.
    b. Your job is to choose projection angles to view the image from. Enter an angle in the '1D proj. angle" space and press "IMAGE 1D" to get your first projection.
    c. Your first projection is now visible in the middle panel, on the first tab. Keep changing the angle and pressing "IMAGE 1D" to get up to five projections.
    d. Using all these projections, decide which of the options is the correct image (which one is consistent with all the projections you got?). Then choose one of the options under "Answer" and press "SUBMIT" to check your answer.
    e. Whether you got it right or wrong, you can press "NEW MODEL" to proceed to the next mystery image!

2. 3D puzzle
    a. Select "3D" on the left panel to enter 3D puzzle mode. A mystery object shows up! The "options" panel on the right shows multiple possible objects to choose from.
    b. Your job is to choose projection angles to view the image from. Choose among "x", "y" and "z" in the '2D proj. axis" field and press "IMAGE 2D" to get your first projection.
    c. Your first projection is now visible in the middle panel, on the first tab. You can now choose one more different axis to get up to two projections.
    d. Using the 1-2 projections, decide which of the options is the correct object (which one is consistent with all the projections you got?). Then choose one of the options under "Answer" and press "SUBMIT" to check your answer.
    e. Whether you got it right or wrong, you can press "NEW MODEL" to proceed to the next mystery object!
3. Hints
    a. The angles and axes directions correspond to the ones in Game 7.
    b. Look over your choices before deciding which angles/axes can best tell them apart.
    c. Use "RESET" to keep the same image/object but get another chance at answering so you can find out the correct answer. If you are having trouble, go through a few questions in this way to understand why the projections look the way they are before proceeding without the help of "RESET".

**Questions**
1. Which of the following is a pair of forward and inverse processes?
    a. Baking a cake and eating it
    b. Growing a tree from a seed and harvesting a seed from that tree
    c. Making a piece of art and selling it to buy more paint
    d. Putting together a LEGO house and taking it apart
2. Which of the following is not a way to sample k-space in a non-cartesian manner?
    a. Sample points with the same spacing across x and across y
    b. Sample points regularly along a line that crosses the center of k-space, rotate the line, and then sample it again
    c. Sample points randomly so more points are sampled from the center of k-space
    d. Sample points along a spiral
3. Which of the following is not a way to process and reconstruct raw MRI data?
    a. Performing an IFFT on it
    b. Convert 1D radial data lines to the spatial domain and perform backprojection
    c. Use gridding and then perform an IFFT
    d. Bake it at 450F for 40 minutes